# Two Wrongs Can Make a Right: A Transfer Learning Approach for Chemical Discovery with Chemical Accuracy

Chenru Duan[1,2], Daniel B. K. Chu[1], Aditya Nandy[1,2], and Heather J. Kulik[1,*]

[1]*Department of Chemical Engineering, Massachusetts Institute of Technology, Cambridge, MA 02139*

[2]*Department of Chemistry, Massachusetts Institute of Technology, Cambridge, MA 02139*

ABSTRACT: Appropriately identifying and treating molecules and materials with significant multi-reference (MR) character is crucial for achieving high data fidelity in virtual high throughput screening (VHTS). Nevertheless, most VHTS is carried out with approximate density functional theory (DFT) using a single functional. Despite development of numerous MR diagnostics, the extent to which a single value of such a diagnostic indicates MR effect on chemical property prediction is not well established. We evaluate MR diagnostics of over 10,000 transition metal complexes (TMCs) and compare to those in organic molecules. We reveal that only some MR diagnostics are transferable across these materials spaces. By studying the influence of MR character on chemical properties (i.e., MR effect) that involves multiple potential energy surfaces (i.e., adiabatic spin splitting, $\Delta E_{\text{H-L}}$, and ionization potential, IP), we observe that cancellation in MR effect outweighs accumulation. Differences in MR character are more important than the total degree of MR character in predicting MR effect in property prediction. Motivated by this observation, we build transfer learning models to directly predict CCSD(T)-level adiabatic $\Delta E_{\text{H-L}}$ and IP from lower levels of theory. By combining these models with uncertainty quantification and multi-level modeling, we introduce a multi-pronged strategy that accelerates data acquisition by at least a factor of three while achieving chemical accuracy (i.e., 1 kcal/mol) for robust VHTS.

# 1. Introduction

Approximate density functional theory (DFT) has become an indispensable workhorse in virtual high throughput screening (VHTS)[1-8] and machine learning (ML)-accelerated chemical discovery[9-12] due to its balanced trade-off in computational cost and accuracy. However, DFT can fail prominently for many of the most promising VHTS targets (e.g., open-shell radicals, transition-metal-containing systems, and strained bonds in transition states).[13-17] These systems may contain strong multi-reference (MR) character due to the existence of near-degenerate orbitals,[18] which cannot be accurately accounted for in DFT due to its single-reference (SR) description of the wavefunction (i.e., the non-interacting system). Although benchmarking studies[19] can be used to identify the best density functional approximation (DFA) that yields accurate energetic properties for specific systems, the choice of DFA varies significantly depending on the systems of interest and cannot be determined *a priori* in VHTS where most materials have yet to be characterized.[20,21] Moreover, an imbalanced treatment of systems that have alternately weak or strong MR character can be expected to undermine the data fidelity and bias the candidate materials uncovered in chemical discovery.[22]

To quantify the degree of MR character, researchers have devised many MR diagnostics[18,23-33] based on different properties (e.g., occupations or atomization energies) and levels of theory. These MR diagnostics can often disagree with each other,[18,34] with the diagnostics derived from DFT being less predictive than those derived from wavefunction theory (WFT).[35] Data-driven methods have augmented conventional approaches[36-40] in MR character classification and method selection. We recently introduced a semi-supervised learning approach to make MR/SR classification based on the consensus of 15 MR diagnostics, which outperforms the traditional cutoff-based approach (i.e., from a single diagnostic) and is transferable to systems of larger sizes and unseen chemical composition.[41] Jeong *et al.*[42] demonstrated an ML protocol that performs an automated selection of active spaces for chemical bond dissociation calculations of main group diatomic molecules, alleviating the computational cost for active space selection.

However, there are still challenges in VHTS arisen from potentially strong MR character. First, the general applicability of MR diagnostics on both organic and transition-metal-containing systems remains unknown since most studies focus solely on organic systems. A notable exception to this is work from Wilson and co-workers[43,44] that demonstrated coupled cluster(CC)-based diagnostics require larger cutoffs on transition metal complexes (TMCs). In addition, while most studies focus on the MR character of a single structure, most chemical properties of interest involve multiple structures and electronic states. How MR character of multiple structures influences the property prediction (i.e., MR effect[18,45]) is not well understood. Despite the development of MR diagnostics and tools for method selection[46], these tools have not been adapted to be suitable for directly improving data quality in VHTS.

In this work, we demonstrate the lack of transferability of MR diagnostics across chemical spaces. We identify the most robust diagnostics to show that imbalances in MR are more important than cumulative MR in properties that depend on multiple electronic states (e.g., adiabatic spin splitting or ionization potential). Motivated by these observations, we train transfer learning models to predict CCSD(T)-level properties from inputs including lower levels of theory (i.e., DFT) and MR diagnostics to reduce DFT errors. We further introduce uncertainty quantification and multi-level modeling into our workflow to only apply the transfer learning predictions when the model has high confidence, accelerating data while achieving chemical accuracy (i.e., within 1 kcal/mol of CCSD(T)) in VHTS.

## 2. Results and Discussion.

### 2a. Limits of MR Diagnostic Transferability.

To evaluate trends in MR character for the 10,000 model TMCs, we used the percentage of correlation energy recovered by CCSD relative to CCSD(T), i.e., $\%E_{corr}[(T)]$, as a figure of merit for measuring the MR character of a system.[35] A smaller $\%E_{corr}[(T)]$ suggests a stronger MR character because CCSD is insufficient to recover the correlation energy. We previously showed[35] that $\%E_{corr}[(T)]$ is system size insensitive and correlates well with $\%E_{corr}[T]$ (i.e., from comparison to full CCSDT, see Computational Details and ESI Figure S1).

Over our data set consisting of low (LS), intermediate (IS), and high spin (HS) complexes, we observe a trend of decreasing MR character with increasing number of unpaired electrons (i.e., LS > IS > HS) (Figure 1). This observation is consistent with both expectations and our prior work[47] and is due to the increased number of accessible configuration state functions in the LS state. Complexes with stronger field ligands (i.e. CO) generally exhibit stronger MR character (Figure 1). For example, we observe decreasing $\%E_{corr}[(T)]$ for complexes with increasing ligand field strength from $H_2O$ to $NH_3$ to CO (Figure 1). This increased MR character can be attributed to the more covalent metal-organic bonding character for complexes with stronger ligand fields. Consequently, when we substitute a $2p$ metal-coordinating with a $3p$ element from the same group (e.g., $NH_3$ to $PH_3$), both the ligand field strength and the MR character of the complexes increases (Figure 1 and ESI Figure S2). In prior work,[48] Feldt *et al.* used increased metal-helium bond lengths to weaken effective ligand fields. Here, we indeed find that the effective ligand fields decrease as the metal-helium bond lengths increase (ESI Figure S3). Concomitant with decreases in effective ligand field as the M–He bond is elongated, we observe decreases in MR character (Figure 1). This trend is in agreement with the observations from spectrochemical series ligands.

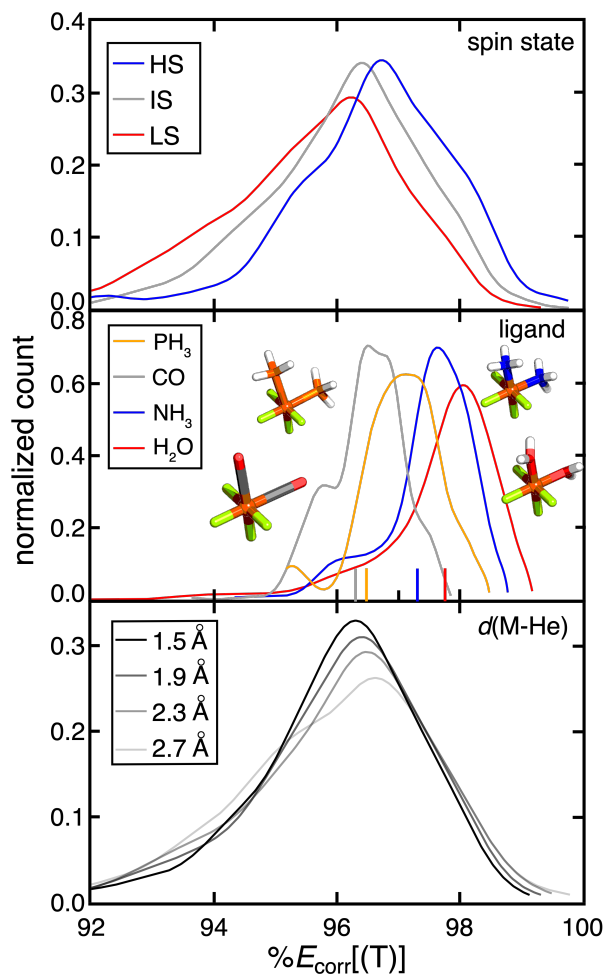

**Figure 1.** Distribution of %$E_{corr}$[(T)] for the 10,000 TMCs categorized by their spin states (top, blue for high spin, gray for intermediate spin, and red for low spin), ligands (middle, orange for PH$_3$, gray for CO, blue for NH$_3$, and red for H$_2$O), and metal-helium distances ($d$(M-He), bottom, with decreasing opacity from 1.0 to 0.4 opacity by increasing $d$(M-He) from 1.5 Å to 2.7 Å). Four representative TMCs with LS Fe(II) and 1.9 Å $d$(M-He) with different ligands in a *cis* configuration are shown. Their corresponding %$E_{corr}$[(T)] values are shown with a colored tick on the x-axis. All atoms are colored as follows: brown for Fe, gray for C, blue for N, red for O, orange for P, white for H, and green for He.

Next, we investigated the linear correlations between pairs of MR diagnostics (ESI Table S1). Consistent with our prior observations on equilibrium and distorted organic molecules,[35] the Pearson correlation coefficients are generally low between pairs of diagnostics obtained from different levels of theory (ESI Figures S4–S5). As was also observed for organic molecules,[35] WFT-based MR diagnostics have better linear and rank-order correlations with $\%E_{corr}[(T)]$ compared to those derived from DFT (ESI Figures S6–S7). This suggests that WFT-based diagnostics are more predictive of whether a system has strong MR character. Nevertheless, fractional occupation-based diagnostics (i.e., Matito's degree of nondynamical correlation, $I_{ND}[B3LYP]$[31,32], and the ratio of nondynamical to total correlation, $r_{ND}[B3LYP]$[34]) are readily obtained at DFT-cost. These low-cost diagnostics have been demonstrated to identify "DFT-safe" islands in VHTS (ESI Table S1).[47] Indeed, $I_{ND}[B3LYP]$[31,32] and $r_{ND}[B3LYP]$[34] yield the best linear and rank-order correlation with $\%E_{corr}[(T)]$ out of the six DFT diagnostics we evaluate (ESI Figures S6–S7). This motivates DFT-based diagnostics in VHTS[47] where MR detection at low cost is essential to avoid computational bottlenecks.

A closely related question is to which extent the MR diagnostics are transferable across different chemical spaces. We compare the relationship between $\%E_{corr}[(T)]$ and representative MR diagnostics for TMCs and equilibrium or stretched organic molecules. We indeed observe divergent behavior for MR diagnostics across these two sets. Organic molecules and TMCs have distinct $T_1$ diagnostic vs. $\%E_{corr}[(T)]$ distributions (Figure 2). This observation highlights that the $T_1$ diagnostic is not a transferable metric for measuring MR character, since organic and TMCs can have different ranges of $T_1$ diagnostic for the same value of $\%E_{corr}[(T)]$. This lack of transferability across materials supports previous arguments for distinct cutoff value choices for the $T_1$ diagnostic when applying it either to organic molecules or inorganic complexes.[43] This non-overlapping 2D distribution for organic molecules and TMCs is indeed observed for all DFT- and CC-based diagnostics and $\%E_{corr}[(T)]$ (ESI Figure S8).

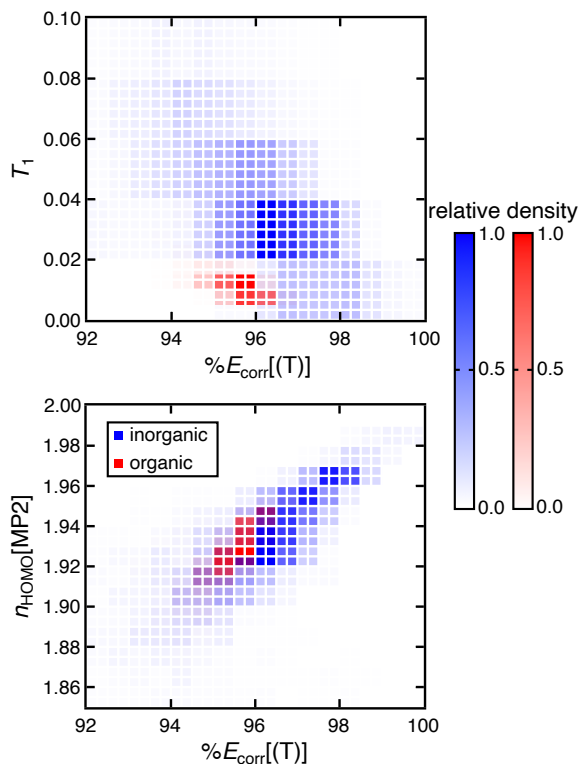

**Figure 2.** 2D histogram for %$E_{corr}$[(T)] vs. $T_1$ (top) and %$E_{corr}$[(T)] vs. $n_{HOMO}$[MP2] (bottom) for the 10,000 TMCs in this work (blue) and for the 12,500 equilibrium or stretched organic molecules in our prior work[35] (red). The relative density of systems within a specific bin is represented by the opacity of the coloring.

The one low-cost diagnostic for which organic molecules and TMCs have overlapping values at the same %$E_{corr}$[(T)] is the MP2-based $n_{HOMO}$[MP2] diagnostic. Since the $n_{HOMO}$[MP2] evaluation is not overly computationally demanding, this analysis highlights its use as a low-cost and transferable metric for MR character determination that could be used across different chemical spaces (Figure 2). Overall, the MP2- and CASSCF-based diagnostics all have a greater degree of overlap with respect to %$E_{corr}$[(T)] for organic molecules and TMCs, suggesting their greater transferability (ESI Figure S9). Surprisingly, the DFT-based $I_{ND}$ and $r_{ND}$ diagnostics, although also motivated from the occupation of virtual orbitals upon electron excitation, are not transferable across chemical spaces (ESI Figure S8). The lack of transferability of $I_{ND}$ and $r_{ND}$ diagnostics could be ascribed to the different degrees of accuracy of DFT on organic molecules and TMCs. One might assume this difference could be attributed to different sizes of organic molecules and TMCs in our set. However, we find invariant 2D distributions of %$E_{corr}$[(T)] vs. MR diagnostics within subsets of organic molecules and TMCs grouped by size, suggesting that this distinct behavior of diagnostics across the two types of molecules is not due to their difference in size (ESI Figure S10–11).

To bridge the gap in performance between low-cost DFT-based diagnostics and computationally demanding WFT-based diagnostics, we trained ANN models to predict the WFT-based diagnostics and %$E_{corr}$[(T)] using DFT-based diagnostics and Coulomb-decay revised autocorrelations (CD-RACs)[35], a set of graph-based descriptors that encode 3D geometric information of TMCs as inputs (see Computational Details). With this approach, we predict WFT-based diagnostics for TMCs with similar accuracies to predictions on organic molecules,[35] despite the poor linear correlations between DFT- and WFT-based diagnostics (Figure 3, ESI Figures S4–5 and Table S2). In addition, we predict %$E_{corr}$[(T)] particularly well solely from DFT-based diagnostics and CD-RACs with a Pearson's $r$ of 0.95 and an MAE of 0.21% (i.e., scaled MAE = 0.015). Given the relatively poor linear correlation between all MR diagnostics and %$E_{corr}$[(T)] for the TMCs, the accurate prediction of %$E_{corr}$[(T)] highlights the utility of our model in practical VHTS (ESI Figure S6).

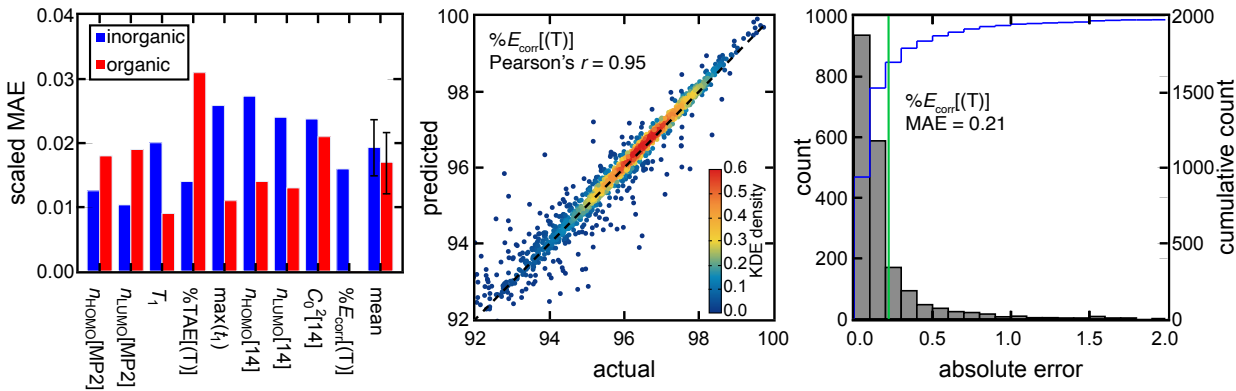

**Figure 3.** (left) Scaled MAE for WFT-based diagnostics and %$E_{corr}$[(T)] for the TMCs (blue) and organic structures (red) on the set-aside test data. The mean scaled MAE for all WFT-based

diagnostics and %$E_{corr}$[(T)] is also shown, with the error bar representing a standard deviation. The scaled MAE is not shown for %$E_{corr}$[(T)] on the organic space, since %$E_{corr}$[(T)] is not an ML model target property in Ref. [35]. (middle) Predicted vs. actual %$E_{corr}$[(T)] on the side-aside test data points of 10,000 TMCs colored by kernel density estimation (KDE) density values, as indicated by inset color bars. A black dashed parity line is also shown. (right) Distributions of absolute test errors for %$E_{corr}$[(T)] (unitless, bins of 0.1) with the MAE annotated as green vertical bars and the cumulative count shown in blue according to the axis on the right.

**2b. Cancellation of Error in MR Effect.**

Numerous efforts[18,29,31,32,34,43] have focused on quantifying the MR character of a single structure and the MR effect on energy evaluation using single-reference methods. However, most property predictions in chemistry are determined from the relative energy of multiple geometric or electronic structures, potentially leading to cancellation of error. Here, we investigate whether the MR effect between multiple structures tends to accumulate or cancel for representative properties. We studied the adiabatic HS to LS splitting, $\Delta E_{H-L}$, which we obtain from the relative electronic energies of two spin states of the same compound in their respective ground state geometries. We also compute the adiabatic ionization potential, IP, which we compute as the electronic energy difference between a molecule before and after electron removal including any reorganization of the oxidized species.

For both properties, we observe that differences in MR character are more important than the total degree of MR character because the MR character cancels when calculating properties involving multiple structures. The error for $\Delta E_{H-L}$ obtained with CCSD in comparison to CCSD(T), i.e., $|\Delta\Delta E_{H-L}[\text{CCSD-CCSD(T)}]|$, correlates well (Pearson's $r$=0.92) with the absolute difference of %$E_{corr}$[(T)] of the two structures (Figure 4). If we instead attempt to predict CCSD errors from the total MR character summed over both structures, we obtain a much poorer correlation (Pearson's $r$=-0.52, Figure 4). To probe why high MR character does not lead to high

MR effect, we considered representative compounds. For the example of *cis* $Cr(II)(NS^-)_2He_4$, the $\Delta\Delta E_{H-L}[CCSD-CCSD(T)]$ is small at ca. 7.2 kcal/mol, although both the LS and HS structures have significant and comparable MR character (LS $\%E_{corr}[(T)]$=89.2, HS $\%E_{corr}[(T)]$=91.1). Another complex, $Co(III)(NH_2^-)He_5$, has a relatively small amount of MR character, as judged by the sum of two spin states (LS $\%E_{corr}[(T)]$=94.9, HS $\%E_{corr}[(T)]$=98.4). However, the imbalance in MR character (i.e., $\%E_{corr}[(T)]$) for the two spin states leads to a large $\Delta\Delta E_{H-L}[CCSD-CCSD(T)]$ (ca. 18.2 kcal/mol). These observations suggest that strong but comparable MR character leads to cancellation of errors in $\Delta E_{H-L}$ for this complex.

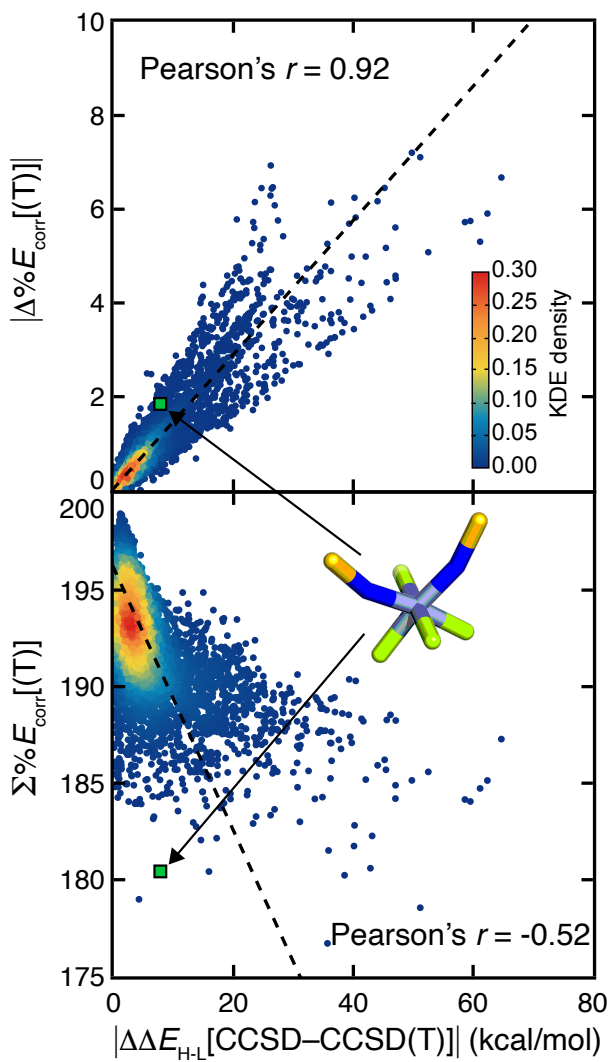

**Figure 4.** The absolute difference in adiabatic spin splitting energy between CCSD and CCSD(T), i.e., $|\Delta\Delta E_{H-L}[\text{CCSD-CCSD(T)}]|$, vs. the absolute difference (top) and the sum (bottom) of $\%E_{corr}[(T)]$ of the two spin states. Points are colored by kernel density estimation (KDE) density values, as indicated by the inset color bar. A black dashed best fit line is shown along with the Pearson correlation coefficient. Cr(II)(NS$^-$)$_2$He$_4$ is shown as a representative example for the cancellation of MR character in property prediction MR effect. Atoms are colored as follows: purple for Cr, gray for C, blue for N, yellow for S, and green for He.

The strong relationship of differences in MR character to errors indicative of MR effect also applies to DFT errors. Choosing B3LYP as a representative functional, we can compare its error in predicting the CCSD(T) $\Delta E_{H-L}$, i.e., $|\Delta\Delta E_{H-L}[\text{B3LYP-CCSD(T)}]|$. We observe a moderate correlation of $|\Delta\Delta E_{H-L}[\text{B3LYP-CCSD(T)}]|$ with the absolute difference of $\%E_{corr}[(T)]$ (Pearson's $r$=0.45), which is stronger than that observed for the sum of $\%E_{corr}[(T)]$ (Pearson's $r$=-0.11) (ESI Figure S12). Observations of better correlations of property errors to MR character differences than to total MR character also hold when evaluating adiabatic IP (ESI Figure S13).

**2c. Transfer Learning to Improve Prediction Accuracy.**

Since high MR character in one structure or electronic state does not necessarily lead to large DFT (or SR-WFT) errors for property evaluations, strategies are needed to predict and correct errors directly on the property of interest. We previously developed an approach[35] to predict the degree of MR character of a single structure at low cost (i.e., with DFT-level diagnostics and CD-RAC descriptors), which we now extend to property prediction (ESI Table S3). Here, we demonstrate a transfer learning approach with ANN models to directly predict the CCSD(T) adiabatic $\Delta E_{H-L}$ and IP from CD-RACs and information obtained from DFT calculations, including the sums and differences of the six DFT-based MR diagnostics and DFT-evaluated $\Delta E_{H-L}$ and IP from four density functionals used in evaluating the six MR diagnostics (i.e., BLYP, B3LYP, PBE, and PBE0, see Methods and ESI Table S3). These trained ANN transfer learning models accurately predict the CCSD(T) result at DFT cost (Figure 5 and ESI

Figure S14). With this model, we obtain a mean absolute error (MAE) of 2.8 kcal/mol for $\Delta E_{\text{H-L}}$ that is three-fold lower than the error obtained from using the B3LYP hybrid functional (Figure 5). We observe similar behavior for the IP, where the transfer learning MAE of 0.14 eV is one third of the error obtained using the B3LYP hybrid functional (ESI Figure S14).

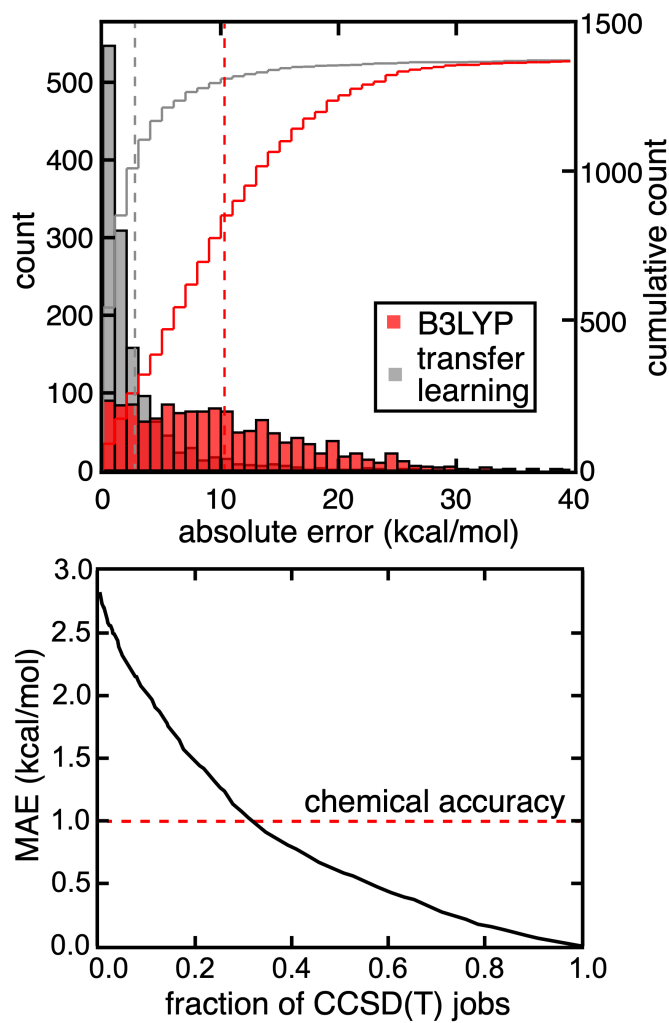

**Figure 5.** Distributions of absolute errors for $\Delta E_{\text{H-L}}$ predicted DFT using B3LYP (red) and transfer learning models (gray) on the set-aside test data, with the cumulative count shown according to the axis on the right (top). The MAEs are shown as vertical bars at 10.2 kcal/mol for DFT and 2.8 kcal/mol for transfer learning. The MAE of the multi-pronged strategy of transfer learning, uncertainty quantification, and multi-theory modeling vs. the percentage of CCSD(T) calculations required (bottom). In all cases, the CCSD(T) results is treated as the reference against which MAEs are evaluated.

In addition, our transfer learning ANN model can be systematically improved by WFT-based (i.e., MP2, CCSD) diagnostics that are more predictive of strong correlation but still lower cost to compute than CCSD(T) (ESI Table S4). For example, by including MP2-based diagnostics (i.e., $n_{HOMO}$[MP2] and $n_{LUMO}$[MP2]) and $\Delta E_{H-L}$ (or IP) computed by MP2, we lower the MAE of $\Delta E_{H-L}$ to 2.2 kcal/mol and IP to 0.12 eV. These MAEs are further reduced to 0.4 kcal/mol for $\Delta E_{H-L}$ and 0.06 eV for IP if we include CCSD-based diagnostics and CCSD-computed $\Delta E_{H-L}$ and IP. More interestingly, we see these large improvements of the transfer learning model performance even though the MP2- or CCSD-evaluated $\Delta E_{H-L}$ and IP do not show significant improvements over DFT in comparison to the CCSD(T) reference (ESI Table S4). This observation suggests our transfer learning models do learn from these WFT-based diagnostics to better predict $\Delta E_{H-L}$ and IP computed by CCSD(T). In addition, we achieve comparable performance to other transfer learning approaches[49,50] demonstrated on organic molecules in terms of scaled MAEs on set-aside test data. One main advance over prior work is that we focus on properties involving multiple electronic states, whereas prior transfer learning demonstrations had been limited to properties determined by a single electronic state and structure (e.g., correlation energy).

Despite the good performance of the transfer learning models to predict the CCSD(T)-level properties, we next investigated whether we could identify compounds with large model uncertainty where errors might also be expected to be large. In such cases, a transfer learning correction may still lead to large errors that would motivate carrying out the full CCSD(T) calculation instead of relying on the transfer learning model. To quantify uncertainty, we used the distance in latent space developed in our previous work[51] as an uncertainty quantification (UQ) metric on our transfer learning models to select complexes that require CCSD(T) calculations. As a demonstration, we selectively perform CCSD(T) calculations on the TMCs that have the largest distance in latent space, and thus highest model uncertainty, but use transfer learning predictions for the others at DFT cost. If we carry out CCSD(T) on 30% of the highest uncertainty points, we reduce errors by a factor of three and achieve chemical accuracy (i.e., 1 kcal/mol) for the prediction of the CCSD(T) $\Delta E_{H-L}$ value (Figure 5). Given that the computational cost of the DFT calculations is negligible relative to CCSD(T) calculations for the moderately-sized TMCs in our dataset, we achieved a three-fold acceleration in data acquisition compared to an all-CCSD(T) approach while maintaining close-to CCSD(T) accuracy. If we aim for an MAE of 1.5 kcal/mol and accept transfer learning predictions on more uncertain points, we reduce the number of complexes that require WFT calculations to only 19% (i.e., five-fold speedup).

Similar speedups were observed for the prediction of adiabatic IP. We only need to carry out 40% of the CCSD(T) calculations with largest ML model uncertainty to achieve a MAE of 0.042 eV (i.e., 1 kcal/mol). The percentage of CCSD(T) calculations carried out can be further reduced to 30% if we aim for a MAE of 0.065 eV (i.e., 1.5 kcal/mol, ESI Figure S14). This strategy of still performing CCSD(T) calculations on large uncertainty points shows significant improvement over the previous strategy[51] where we would instead just avoid making a prediction on the large uncertainty points. For example, we must discard 74% of complexes for $\Delta E_{H-L}$ and 90% of the complexes for IP transfer learning to retain the chemical accuracy of 1 kcal/mol on the points for which a prediction is still made (ESI Figure S15). Thus, we have introduced a multi-pronged strategy of transfer learning, ML model UQ, and multi-level modeling to accelerate data acquisition while maintaining high overall data fidelity for chemical discovery.

**3. Conclusions.**

In conclusion, we studied trends in MR character of over 10,000 TMCs. Over this set, we observed that complexes with more unpaired electrons (i.e., LS) and stronger ligand fields have more significant MR character. Taking both organic molecules and TMCs into consideration, we showed that DFT- and CC-based diagnostics (e.g., $T_1$ diagnostic) have distinct relationships with %$E_{corr}$[(T)] for the two classes of molecules, thus limiting their transferability. In contrast, MP2- and CASSCF-based diagnostics have more consistent relationships with %$E_{corr}$[(T)] for both organic molecules and TMCs, demonstrating greater transferability. Therefore, we built ML models to predict these computationally demanding and transferable WFT-based diagnostics and %$E_{corr}$[(T)] from less costly DFT-based diagnostics. We obtained excellent accuracy to directly predict %$E_{corr}$[(T)] (i.e., MAE = 0.21%), motivating the use of our models in VHTS.

Motivated by the fact that most chemical properties are determined from the relative energy of multiple geometric or electronic structures, we next investigated the MR effect on two properties dependent on multiple optimized geometries, the adiabatic spin splitting ($\Delta E_{H-L}$) and IP. We observed that differences in MR character are more important than the total degree of MR character, suggesting that cancellation in MR effect outweighs the accumulation. As a result, strong MR character in a single structure does not necessarily lead to large DFT errors. Motivated by this observation, we built two ML models to directly predict the CCSD(T) adiabatic $\Delta E_{H-L}$ and IP *via* a transfer learning approach. This approach demonstrated a three-fold reduction in errors compared to using B3LYP on both properties. Finally, we introduced UQ and multi-level modeling into our workflow in which we carried out CCSD(T) calculations on most uncertain points but used transfer learning predictions on the others. We demonstrated that this multi-pronged strategy accelerates data acquisition by a factor of three while maintaining high overall data fidelity (i.e., 1 kcal/mol chemical accuracy) for chemical discovery. We anticipate our observations on the cancellation of MR effects in property evaluations and our multi-pronged strategy to overcome cost-accuracy trade-off limitations in VHTS for challenging materials spaces.

### 4. Computational Details.

*Data Sets*. Mononuclear octahedral transition metal complexes (TMCs) with Cr, Mn, Fe, and Co in +2 and +3 oxidation states were studied in up to three spin states, i.e., high, intermediate, and low, as follows: quintet, triplet, and singlet for $d^6$ Co(III)/Fe(II) and $d^4$ Mn(III)/Cr(II); sextet, quartet, and doublet for $d^5$ Fe(III)/Mn(II), and quartet and doublet for $d^3$ Cr(III) and $d^7$ Co(II) (ESI Table S5). We used monodentate ligands from both the spectrochemical series[52] and our prior OHLDB set[53] (ESI Table S6). To restrict the system size, we employed He atoms as ligands for four to six of the six ligands. For the remaining one to two non-He ligands, we considered both *cis* and *trans* symmetry, and we varied the metal-He distance to mimic ligand field strength differences while all other metal-ligand distances were freely optimized (Figure 1 and ESI Table S7).

*DFT geometry optimizations*. DFT geometry optimizations with the B3LYP[54-56] global hybrid functional were carried out using a developer version of graphical-processing unit (GPU)-accelerated electronic structure code TeraChem[57,58] The LANL2DZ effective core potential[59] basis set was used for metals and the 6-31G* basis for all other atoms. Singlet spin states were calculated with the spin-restricted formalism while all other calculations were carried out in a spin-unrestricted formalism. In all DFT geometry optimizations, level shifting[60] of 0.25 Ha on majority virtual spin orbitals and 0.25 Ha on minority virtual spin orbitals was employed. Initial geometries were assembled by molSimplify[61,62] and optimized using the L-BFGS algorithm in translation rotation internal coordinates (TRIC)[63] to the default tolerances of $4.5 \times 10^{-4}$ hartree/bohr for the maximum gradient and $1 \times 10^{-6}$ hartree for the energy change between steps. During the optimization, the positions of the metal and He atoms were fixed to maintain the target metal-He distances and angles. Geometry checks[64,65] were applied to eliminate optimized structures that deviated from the expected octahedral shape following previously established metrics[64,65] without modification (ESI Table S7).

*MR diagnostic calculations.* Following our prior studies[35,41], we calculated 14 MR diagnostics[18,23-32] using ORCA 4.0.2.1[66,67] with the cc-pVTZ basis set on the metals as well as P and S elements and the cc-pVDZ basis set on all other atoms (ESI Table S1). To evaluate the MR character, the restricted open-shell formalism was used in all DFT and Hartree-Fock (HF) calculations. We chose the restricted open-shell formalism since it was observed[18,43] that unrestricted formalism can recover some MR effects in open-shell systems and thus lead to smaller MR diagnostics. We converged a B3LYP calculation and used it to initialize both DFT calculations with other density functionals (i.e., BLYP, B1LYP, PBE, and PBE0) and the HF calculations. This ensured we converged a consistent electronic state over multiple calculations and also saved computational time. The converged HF wavefunction was then used for MP2 and CCSD(T) calculations. Finally, the MP2 natural orbitals were used to set up a CASSCF calculation with active spaces of 10, 12, and 14 orbitals (ESI Figure S16).

All MR diagnostics were computed using the default parameters in ORCA (ESI Table S8). During the computation of total atomization energy (TAE) based diagnostics, we assumed heterolytic dissociation for the metal-ligand bond (i.e., the oxidation state of the metal does not change) and homolytic dissociation for the atoms in the ligands, where each individual atom kept its formal charge (ESI Table S6). We chose the percentage of correlation energy recovered by CCSD compared to CCSD(T) (i.e., %$E_{corr}$[(T)]) as the figure of merit as we observed good correspondence of %$E_{corr}$[(T)] and %$E_{corr}$[T] in both equilibrium and distorted organic molecule in our previous work[35]. We also tested it for complexes with six helium atoms as ligands in this work and found excellent agreement between %$E_{corr}$[(T)] and %$E_{corr}$[T] (ESI Figure S4)

If all 14 MR diagnostics could not be successfully computed (e.g., due to lack of SCF convergence or one of the calculations exceeding the allowed wall time), we removed the TMC from the dataset (ESI Table S9). A few (274, ca. 2%) CCSD(T) calculations resulted in significantly different perturbative triple corrections among TMCs with different metal–He distances but same chemical composition, potentially due to the CCSD wavefunctions converging to different electronic states. We removed those cases by the Grubbs outlier test[68] and Z-score test by comparing the perturbative triple corrections obtained using TMCs with the same chemical composition and ligand symmetry but different metal–He distances (ESI Figure S17 and Tables S10–S11). We also removed TMCs where the standard deviation of the leading weight of the CASSCF wavefunction, $C_0^2$, obtained by the three active spaces (i.e., with 10, 12, and 14 active orbitals) was larger than 0.1 (334, ca. 3%), which indicated that a final active space of 14 orbitals was not sufficient (ESI Figure S18 and Table S12).

*ML models*. As in prior work[35,41], we use Coulomb-decay revised auto-correlations (CD-RACs)[35] as descriptors for all of our machine learning models. CD-RACs are sums of products and differences of five atom-wise heuristic properties (i.e., topology, identity, electronegativity, covalent radius, and nuclear charge) on the 2D molecular graph divided by the pairwise atomic distance. This incorporation of the pairwise distance imparts 3D geometric information to graph-based RACs[69] to distinguish TMCs with the same chemical composition but different metal–He distances. We chose CD-RACs as descriptors because RACs have been previously demonstrated to provide good performance in equilibrium properties of transition metal complexes[65,70] and CD-RACs have shown superior performance on predicting MR diagnostics on both equilibrium and non-equilibrium geometries of organic molecules in comparison to several alternatives.[35] As motivated previously,[69] we apply the maximum bond depth of three and eliminate constant RACs (ESI Text S1). For properties that involve two structures (i.e., adiabatic spin splitting and ionization potential), the CD-RACs of the two geometries were concatenated (ESI Table S3). For all artificial neural network (ANN) models, the hyperparameters were selected using HyperOpt[71] with 200 evaluations, using a random 80/20 train/test split, with 20% of the training data (i.e., 16% overall) used as the validation set (ESI Table S13). All ANN models were trained using Keras[72] with Tensorflow[73] as a backend. All models used the Adam optimizer up to 2,000 epochs, and dropout, batch normalization, and early stopping to avoid over-fitting.

ASSOCIATED CONTENT

**Electronic Supporting Information Statement**. %$E_{corr}$[T] vs. %$E_{corr}$[(T)] for select complexes; $\Delta E_{H-L}$ versus metal-helium distance for Fe(II)(CO)He$_5$; distribution of %$E_{corr}$[(T)]; level of theory for each computed MR diagnostic; Pearson's $r$ matrix of MR diagnostics and %$E_{corr}$[(T)]; bar plot of unsigned Pearson's $r$ and Spearman's $r$ for MR diagnostics; 2D histogram for %$E_{corr}$[(T)] and MR diagnostics; summary of ANN model performance; MR effect cancellation vs. accumulation for adiabatic $\Delta E_{H-L}$ and IP; summary of ML features; MAE of transfer learning models on adiabatic $\Delta E_{H-L}$ and IP; comparison of the combined strategy and UQ only approach; metals, oxidation states, spin states, and ligands for constructing the data set; data attrition counts, reasons, and cutoffs; workflow of computing MR diagnostics and calculation parameters; potential energy curves and $T_1$ diagnostic for select TMCs; extended description of the CD-RAC featurization; range of hyperparameters during hyperparameter selections.

AUTHOR INFORMATION

**Corresponding Author**

*email:hjkulik@mit.edu19

**Notes**

The authors declare no competing financial interest.

ACKNOWLEDGMENT

This work was supported by the Department of Energy under grant number DE-SC0018096 and the Office of Naval Research under grant numbers N00014-18-1-2434 and N00014-20-1-2150. A.N. and D.B.K.C. were partially supported by a National Science Foundation Graduate Research Fellowship under Grant #1122374. H.J.K. holds a Career Award at the Scientific Interface from the Burroughs Wellcome Fund and an AAAS Marion Milligan Mason Award, which supported this work. This work was carried out in part using computational resources from the Extreme Science and Engineering Discovery Environment (XSEDE), which is supported by National Science Foundation grant number ACI-1548562. The authors thank Adam H. Steeves for providing a critical reading of the manuscript.

References

(1) Shu, Y. N.; Levine, B. G. Simulated Evolution of Fluorophores for Light Emitting Diodes. *J. Chem. Phys.* **2015,** *142*, 104104.
(2) Gomez-Bombarelli, R.; Aguilera-Iparraguirre, J.; Hirzel, T. D.; Duvenaud, D.; Maclaurin, D.; Blood-Forsythe, M. A.; Chae, H. S.; Einzinger, M.; Ha, D. G.; Wu, T.et al. Design of Efficient Molecular Organic Light-Emitting Diodes by a High-Throughput Virtual Screening and Experimental Approach. *Nat. Mater.* **2016,** *15*, 1120-+.
(3) Kanal, I. Y.; Owens, S. G.; Bechtel, J. S.; Hutchison, G. R. Efficient Computational Screening of Organic Polymer Photovoltaics. *J. Phys. Chem. Lett.* **2013,** *4*, 1613-1623.
(4) Vogiatzis, K. D.; Polynski, M. V.; Kirkland, J. K.; Townsend, J.; Hashemi, A.; Liu, C.; Pidko, E. A. Computational Approach to Molecular Catalysis by 3d Transition Metals: Challenges and Opportunities. *Chem. Rev.* **2018,** *119*, 2453-2523.
(5) Foscato, M.; Jensen, V. R. Automated in Silico Design of Homogeneous Catalysts. *ACS Catal.* **2020,** *10*, 2354-2377.
(6) Curtarolo, S.; Hart, G. L.; Nardelli, M. B.; Mingo, N.; Sanvito, S.; Levy, O. The High-Throughput Highway to Computational Materials Design. *Nat. Mater.* **2013,** *12*, 191-201.
(7) Ong, S. P.; Richards, W. D.; Jain, A.; Hautier, G.; Kocher, M.; Cholia, S.; Gunter, D.; Chevrier, V. L.; Persson, K. A.; Ceder, G. Python Materials Genomics (Pymatgen): A Robust, Open-Source Python Library for Materials Analysis. *Comput. Mater. Sci.* **2013,** *68*, 314-319.
(8) Nørskov, J. K.; Bligaard, T. The Catalyst Genome. *Angew. Chem. Int. Ed.* **2013,** *52*, 776-777.
(9) Janet, J. P.; Duan, C. R.; Nandy, A.; Liu, F.; Kulik, H. J. Navigating Transition-Metal Chemical Space: Artificial Intelligence for First-Principles Design. *Accounts Chem Res* **2021,** *54*, 532-545.
(10) Rosen, A. S.; Iyer, S. M.; Ray, D.; Yao, Z. P.; Aspuru-Guzik, A.; Gagliardi, L.; Notestein, J. M.; Snurr, R. Q. Machine Learning the Quantum-Chemical Properties of Metal-Organic Frameworks for Accelerated Materials Discovery. *Matter-Us* **2021,** *4*, 1578-1597.
(11) Ceriotti, M.; Clementi, C.; von Lilienfeld, O. A. Introduction: Machine Learning at the Atomic Scale. *Chem. Rev.* **2021,** *121*, 9719-9721.
(12) Keith, J. A.; Vassilev-Galindo, V.; Cheng, B. Q.; Chmiela, S.; Gastegger, M.; Mueller, K. R.; Tkatchenko, A. Combining Machine Learning and Computational Chemistry for Predictive Insights into Chemical Systems. *Chem. Rev.* **2021,** *121*, 9816-9872.
(13) Cohen, A. J.; Mori-Sánchez, P.; Yang, W. Challenges for Density Functional Theory. *Chem. Rev.* **2012,** *112*, 289-320.
(14) Becke, A. D. Perspective: Fifty Years of Density-Functional Theory in Chemical Physics. *J. Chem. Phys.* **2014,** *140*, 18A301.
(15) Cramer, C. J.; Truhlar, D. G. Density Functional Theory for Transition Metals and Transition Metal Chemistry. *Phys. Chem. Chem. Phys.* **2009,** *11*, 10757-10816.
(16) Duan, C.; Liu, F.; Nandy, A.; Kulik, H. J. Putting Density Functional Theory to the Test in Machine-Learning-Accelerated Materials Discovery. *J. Phys. Chem. Lett.* **2021,** *12*, 4628-4637.
(17) Yu, H. Y. S.; Li, S. H. L.; Truhlar, D. G. Perspective: Kohn-Sham Density Functional Theory Descending a Staircase. *J. Chem. Phys.* **2016,** *145*, 130901.
(18) Fogueri, U. R.; Kozuch, S.; Karton, A.; Martin, J. M. L. A Simple Dft-Based Diagnostic for Nondynamical Correlation. *Theoretical Chemistry Accounts* **2013,** *132*, 1291.
(19) Mardirossian, N.; Head-Gordon, M. Thirty Years of Density Functional Theory in Computational Chemistry: An Overview and Extensive Assessment of 200 Density Functionals. *Mol. Phys.* **2017,** *115*, 2315-2372.
(20) Duan, C.; Chen, S. X.; Taylor, M. G.; Liu, F.; Kulik, H. J. Machine Learning to Tame Divergent Density Functional Approximations: A New Path to Consensus Materials Design Principles. *Chem. Sci.* **2021,** *12*, 13021-13036.
(21) Reiher, M. Molecule-Specific Uncertainty Quantification in Quantum Chemical Studies. *Isr. J. Chem.* **2021**, DOI:ARTN e202100101
10.1002/ijch.202100101 ARTN e202100101
10.1002/ijch.202100101.
(22) Nandy, A.; Duan, C.; Kulik, H. J. Audacity of Huge: Overcoming Challenges of Data Scarcity and Data Quality for Machine Learning in Computational Materials Discovery. *Curr. Opin. Chem. Eng. .* **2022,** *in press*.
(23) Lee, T. J.; Taylor, P. R. A Diagnostic for Determining the Quality of Single-Reference Electron Correlation Methods. *Int. J. Quantum Chem.* **1989**, DOI:10.1002/qua.560360824 10.1002/qua.560360824, 199-207.

(24) Sears, J. S.; Sherrill, C. D. Assessing the Performance of Density Functional Theory for the Electronic Structure of Metal-Salens: The D(2)-Metals. *J. Phys. Chem. A* **2008,** *112*, 6741-6752.
(25) Sears, J. S.; Sherrill, C. D. Assessing the Performance of Density Functional Theory for the Electronic Structure of Metal-Salens: The 3d(0)-Metals. *J. Phys. Chem. A* **2008,** *112*, 3466-3477.
(26) Langhoff, S. R.; Davidson, E. R. Configuration Interaction Calculations on the Nitrogen Molecule. *Int. J. Quantum Chem.* **1974,** *8*, 61-72.
(27) Janssen, C. L.; Nielsen, I. M. B. New Diagnostics for Coupled-Cluster and Moller-Plesset Perturbation Theory. *Chem. Phys. Lett.* **1998,** *290*, 423-430.
(28) Nielsen, I. M. B.; Janssen, C. L. Double-Substitution-Based Diagnostics for Coupled-Cluster and Moller-Plesset Perturbation Theory. *Chem. Phys. Lett.* **1999,** *310*, 568-576.
(29) Schultz, N. E.; Zhao, Y.; Truhlar, D. G. Density Functionals for Inorganometallic and Organometallic Chemistry. *J. Phys. Chem. A* **2005,** *109*, 11127-11143.
(30) Tishchenko, O.; Zheng, J. J.; Truhlar, D. G. Multireference Model Chemistries for Thermochemical Kinetics. *J. Chem. Theory Comput.* **2008,** *4*, 1208-1219.
(31) Ramos-Cordoba, E.; Matito, E. Local Descriptors of Dynamic and Nondynamic Correlation. *J. Chem. Theory Comput.* **2017,** *13*, 2705-2711.
(32) Ramos-Cordoba, E.; Salvador, P.; Matito, E. Separation of Dynamic and Nondynamic Correlation. *Phys. Chem. Chem. Phys.* **2016,** *18*, 24015-24023.
(33) Karton, A.; Daon, S.; Martin, J. M. L. W4-11: A High-Confidence Benchmark Dataset for Computational Thermochemistry Derived from First-Principles W4 Data. *Chem. Phys. Lett.* **2011,** *510*, 165-178.
(34) Kesharwani, M. K.; Sylvetsky, N.; Kohn, A.; Tew, D. P.; Martin, J. M. L. Do Ccsd and Approximate Ccsd-F12 Variants Converge to the Same Basis Set Limits? The Case of Atomization Energies. *J. Chem. Phys.* **2018,** *149*, 154109.
(35) Duan, C.; Liu, F.; Nandy, A.; Kulik, H. J. Data-Driven Approaches Can Overcome the Cost– Accuracy Trade-Off in Multireference Diagnostics. *J. Chem. Theory Comput.* **2020,** https://dx.doi.org/10.1021/acs.jctc.0c00358, 4373-4387.
(36) Stein, C. J.; Reiher, M. Automated Selection of Active Orbital Spaces. *J. Chem. Theory Comput.* **2016,** *12*, 1760-1771.
(37) Schriber, J. B.; Evangelista, F. A. Communication: An Adaptive Configuration Interaction Approach for Strongly Correlated Electrons with Tunable Accuracy. *J. Chem. Phys.* **2016,** *144*, 161106.
(38) Baiardi, A.; Kelemen, A. K.; Reiher, M. Excited-State Dmrg Made Simple with Feast. *J. Chem. Theory Comput.* **2021**, DOI:10.1021/acs.jctc.1c00984 10.1021/acs.jctc.1c00984.
(39) King, D. S.; Gagliardi, L. A Ranked-Orbital Approach to Select Active Spaces for High-Throughput Multireference Computation. *J. Chem. Theory Comput.* **2021,** *17*, 2817-2831.
(40) He, N.; Evangelista, F. A. A Zeroth-Order Active-Space Frozen-Orbital Embedding Scheme for Multireference Calculations. *J. Chem. Phys.* **2020,** *152*.
(41) Duan, C.; Liu, F.; Nandy, A.; Kulik, H. J. Semi-Supervised Machine Learning Enables the Robust Detection of Multireference Character at Low Cost. *J. Phys. Chem. Lett.* **2020,** *11*, 6640-6648.
(42) Jeong, W.; Stoneburner, S. J.; King, D.; Li, R.; Walker, A.; Lindh, R.; Gagliardi, L. Automation of Active Space Selection for Multireference Methods Via Machine

Learning on Chemical Bond Dissociation. *J. Chem. Theory Comput.* **2020,** *16*, 2389-2399.
(43) Jiang, W. Y.; DeYonker, N. J.; Wilson, A. K. Multireference Character for 3d Transition-Metal-Containing Molecules. *J. Chem. Theory Comput.* **2012,** *8*, 460-468.
(44) Wang, J.; Manivasagam, S.; Wilson, A. K. Multireference Character for 4d Transition Metal-Containing Molecules. *J. Chem. Theory Comput.* **2015,** *11*, 5865-5872.
(45) Sprague, M. K.; Irikura, K. K. In *Thom H. Dunning, Jr.*; Springer, 2015,307-318
(46) McAnanama-Brereton, S.; Waller, M. P. Rational Density Functional Selection Using Game Theory. *Journal of Chemical Information and Modeling* **2017,** *58*, 61-67.
(47) Liu, F.; Duan, C.; Kulik, H. J. Rapid Detection of Strong Correlation with Machine Learning for Transition-Metal Complex High-Throughput Screening. *J. Phys. Chem. Lett.* **2020,** *11*, 8067-8076.
(48) Feldt, M.; Phung, Q. M.; Pierloot, K.; Mata, R. A.; Harvey, J. N. Limits of Coupled-Cluster Calculations for Non-Heme Iron Complexes. *J. Chem. Theory Comput.* **2019,** *15*, 922-937.
(49) Husch, T.; Sun, J. C.; Cheng, L. X.; Lee, S. J. R.; Miller, T. F. Improved Accuracy and Transferability of Molecular-Orbital-Based Machine Learning: Organics, Transition-Metal Complexes, Non-Covalent Interactions, and Transition States. *J. Chem. Phys.* **2021,** *154*.
(50) Smith, J. S.; Nebgen, B. T.; Zubatyuk, R.; Lubbers, N.; Devereux, C.; Barros, K.; Tretiak, S.; Isayev, O.; Roitberg, A. E. Approaching Coupled Cluster Accuracy with a General-Purpose Neural Network Potential through Transfer Learning. *Nat. Commun.* **2019,** *10*, 2903.
(51) Janet, J. P.; Duan, C.; Yang, T.; Nandy, A.; Kulik, H. J. A Quantitative Uncertainty Metric Controls Error in Neural Network-Driven Chemical Discovery. *Chem. Sci.* **2019**, DOI:10.1039/C9SC02298H 10.1039/C9SC02298H.
(52) Tsuchida, R. Absorption Spectra of Co-Ordination Compounds. I. *BCSJ* **1938,** *13*, 388-400.
(53) Gugler, S.; Janet, J. P.; Kulik, H. J. Enumeration of De Novo Inorganic Complexes for Chemical Discovery and Machine Learning. *Mol. Syst. Des. Eng.* **2020,** *5*, 139-152.
(54) Lee, C.; Yang, W.; Parr, R. G. Development of the Colle-Salvetti Correlation-Energy Formula into a Functional of the Electron Density. *Phys. Rev. B* **1988,** *37*, 785-789.
(55) Becke, A. D. Density‐Functional Thermochemistry. Iii. The Role of Exact Exchange. *J. Chem. Phys.* **1993,** *98*, 5648-5652.
(56) Stephens, P. J.; Devlin, F. J.; Chabalowski, C. F.; Frisch, M. J. Ab Initio Calculation of Vibrational Absorption and Circular Dichroism Spectra Using Density Functional Force Fields. *The Journal of Physical Chemistry* **1994,** *98*, 11623-11627.
(57) Seritan, S.; Bannwarth, C.; Fales, B. S.; Hohenstein, E. G.; Isborn, C. M.; Kokkila-Schumacher, S. I. L.; Li, X.; Liu, F.; Luehr, N.; Snyder, J. W.et al. Terachem: A Graphical Processing Unit-Acceleratedelectronic Structure Package Forlarge-Scaleab Initio Molecular Dynamics. *Wires Comput Mol Sci* **2021,** *11*.
(58) Petachem LLC. Petachem. http://www.petachem.com. (Accessed May 23, 2020).
(59) Hay, P. J.; Wadt, W. R. Ab Initio Effective Core Potentials for Molecular Calculations. Potentials for the Transition Metal Atoms Sc to Hg. *J. Chem. Phys.* **1985,** *82*, 270-283.
(60) Saunders, V. R.; Hillier, I. H. Level-Shifting Method for Converging Closed Shell Hartree-Fock Wave-Functions. *Int. J. Quantum Chem.* **1973,** *7*, 699-705.

(61) Ioannidis, E. I.; Gani, T. Z. H.; Kulik, H. J. Molsimplify: A Toolkit for Automating Discovery in Inorganic Chemistry. *J. Comput. Chem.* **2016,** *37*, 2106-2117.
(62) molSimplify. Molsimplify Github Page. https://github.com/hjkgrp/molSimplify. (Accessed January 21, 2019).
(63) Wang, L.-P.; Song, C. Geometry Optimization Made Simple with Translation and Rotation Coordinates. *J. Chem. Phys.* **2016,** *144*, 214108.
(64) Duan, C.; Janet, J. P.; Liu, F.; Nandy, A.; Kulik, H. J. Learning from Failure: Predicting Electronic Structure Calculation Outcomes with Machine Learning Models. *J. Chem. Theory Comput.* **2019,** *15*, 2331-2345.
(65) Nandy, A.; Duan, C.; Janet, J. P.; Gugler, S.; Kulik, H. J. Strategies and Software for Machine Learning Accelerated Discovery in Transition Metal Chemistry. *Ind. Eng. Chem. Res.* **2018,** *57*, 13973-13986.
(66) Neese, F. The Orca Program System. *Wires Comput Mol Sci* **2012,** *2*, 73-78.
(67) Neese, F. Software Update: The Orca Program System, Version 4.0. *Wires Comput Mol Sci* **2018,** *8*, e1327.
(68) Grubbs, F. E. Sample Criteria for Testing Outlying Observations. *Ann. Math. Stat.* **1950,** *21*, 27-58.
(69) Janet, J. P.; Kulik, H. J. Resolving Transition Metal Chemical Space: Feature Selection for Machine Learning and Structure-Property Relationships. *J. Phys. Chem. A* **2017,** *121*, 8939-8954.
(70) Janet, J. P.; Ramesh, S.; Duan, C.; Kulik, H. J. Accurate Multi-Objective Design in a Space of Millions of Transition Metal Complexes with Neural-Network-Driven Efficient Global Optimization. *ACS Cent. Sci.* **2020,** *6*, 513-524.
(71) Bergstra, J. C., D. D.; Yamins, D. Hyperopt: A Python Library for Optimizing the Hyperparameters of Machine Learning Algorithms *Proceedings of the 12th Python in science conference* **2013**, 13-20.
(72) Chollet, F. Keras. https://keras.io. (Accessed June 24, 2021).
(73) Abadi, M.; Agarwal, A.; Barham, P.; Brevdo, E.; Chen, Z.; Citro, C.; Corrado, G. S.; Davis, A.; Dean, J.; Devin, M.et al., 2015.

Chenru Duan[1,2], Daniel B. K. Chu[1], Aditya Nandy[1,2], and Heather J. Kulik[1,*]

[1]Department of Chemical Engineering, Massachusetts Institute of Technology, Cambridge, MA 02139

[2]Department of Chemistry, Massachusetts Institute of Technology, Cambridge, MA 02139

**Contents**

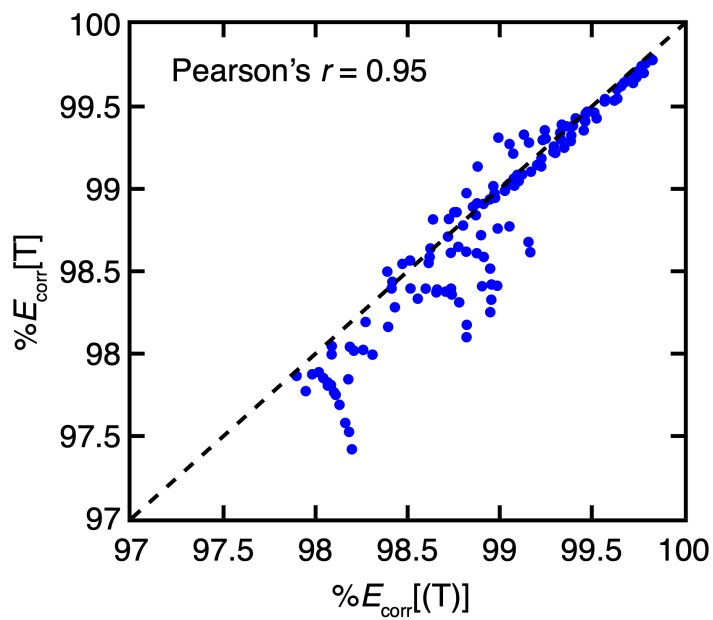

**Figure S1.** %$E_{corr}$[T] vs. %$E_{corr}$[(T)] for 132 complexes with six helium atoms as ligands at various metal-helium bond lengths.

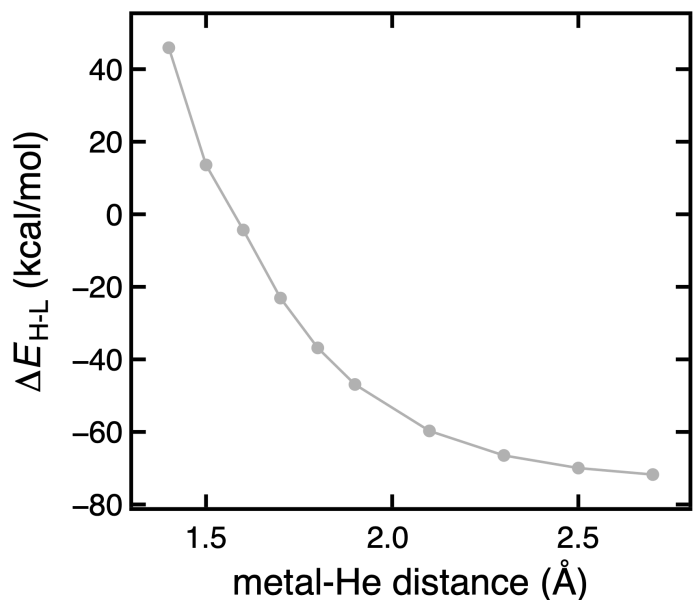

**Figure S2.** $\Delta E_{H-L}$ for Fe(II)(CO)(He)$_5$ versus metal-helium distance in Å. The Fe-C distance of the carbonyl ligand is relaxed freely while the metal-helium distance is constrained. $\Delta E_{H-L}$ were computed with CCSD(T) at a mixed basis set with cc-pVTZ on Fe and cc-pVDZ on C, O, and He.

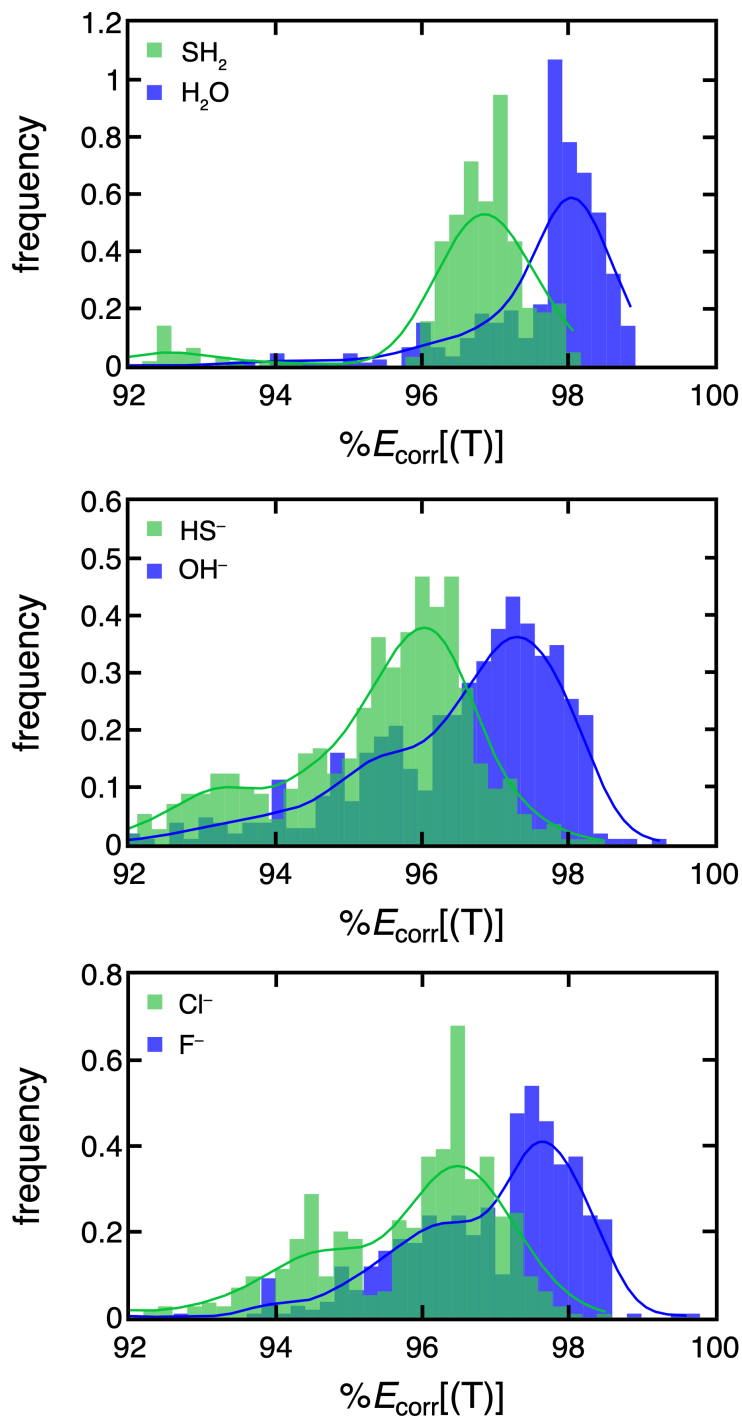

**Figure S3.** Distribution of %$E_{corr}$[(T)] for TMCs with isovalent *2p*- and *3p*-coordinating ligands. (top) H$_2$O (blue) and SH$_2$ (green); (middle) OH$^-$ (blue) and HS$^-$ (green); (bottom) F$^-$ (blue) and Cl$^-$ (green). TMCs with both one and two non He ligands are shown together in the distributions.

**Table S1.** Summary of MR diagnostics grouped by type and method used. Compared to our prior work[1], $D_1$ and $D_2$ diagnostics are removed considering their high linear correlations with some of the diagnostics ($D_1$ and $T_1$, $D_2$ and $C_0^2$).

| Diagnostic | Method | Type | Extended description |
|---|---|---|---|
| $B_1$[2] | DFT | TAE | Differences in total atomization energy for BLYP and B1LYP (25% exchange) divided by number of pairs of bonded atoms |
| $A_{25}[PBE]$[3] | DFT | TAE | 4x the difference in TAE[PBE] and TAE[PBE0] (25% exchange) divided by TAE[PBE] |
| $I_{ND}[PBE]$[4-5] | DFT | occupations | Estimation of non-dynamical contribution from finite-temperature DFT with PBE functional (T = 5000 K) |
| $r_{ND}[PBE]$[6] | DFT | occupations | ratio of FT-DFT $I_{ND}$ from PBE to the sum of $I_{ND}$ with the dynamical term, $I_D$ |
| $I_{ND}[B3LYP]$[4-5] | DFT | occupations | Estimation of non-dynamical contribution from finite-temperature DFT with B3LYP functional (T = 9000 K) |
| $r_{ND}[B3LYP]$[6] | DFT | occupations | ratio of FT-DFT $I_{ND}$ from B3LYP to the sum of $I_{ND}$ with the dynamical term, $I_D$ |
| $n_{HOMO}[MP2]$[3, 7] | MP2 | occupations | MP2 highest occupied natural orbital occupation |
| $n_{LUMO}[MP2]$[3, 7] | MP2 | occupations | MP2 lowest unoccupied natural orbital occupation |
| $T_1$[8] | CCSD | excitations | Frobenius norm of the single-excitation amplitude vector normalized by the square root of the number of electrons in CCSD |
| $\max(t_1)$[9] | CCSD | excitations | The largest eigenvalue of the matrix derived from the single-excitation amplitudes. |
| %TAE[(T)][10] | CCSD(T) | TAE | Percent difference in TAE from CCSD vs. CCSD(T) |
| $C_0^2[14]$[8, 11-13] | CASSCF | occupations | CASSCF leading coefficient CSF at an active space of 14 orbitals |
| $n_{HOMO}[14]$[3, 14] | CASSCF | occupations | CASSCF highest occupied natural orbital occupation at an active space of 14 orbitals |
| $n_{LUMO}[14]$[3, 14] | CASSCF | occupations | CASSCF lowest unoccupied natural orbital occupation at an active space of 14 orbitals |

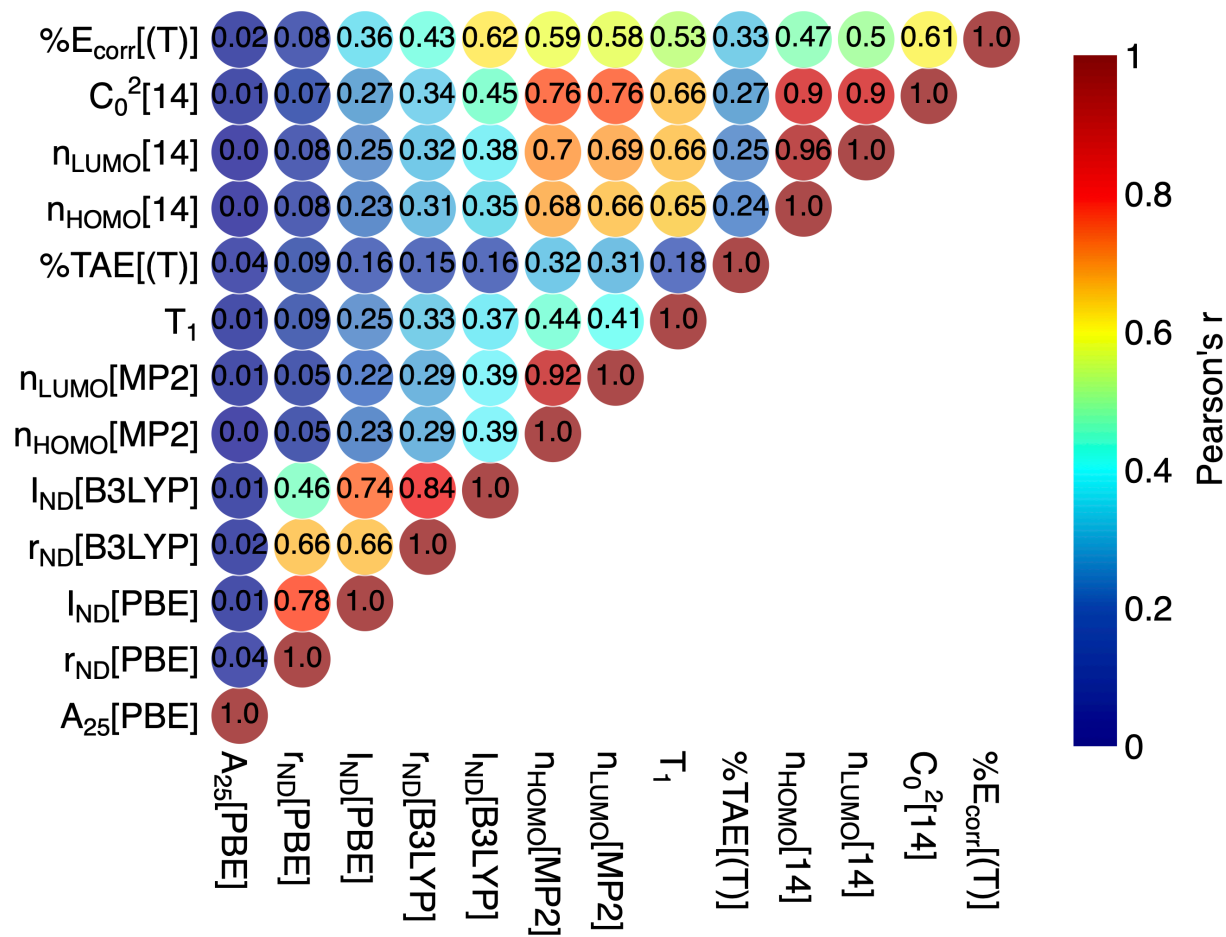

**Figure S4.** An upper triangular matrix of unsigned Pearson's $r$ for pairs of MR diagnostics and %$E_{corr}$[(T)] on the set of 10,000 TMCs. For each pair, it is colored by the unsigned Pearson's $r$ and the $r$ value explicitly shown in the circle. Fourteen active orbitals are used for the CASSCF calculations

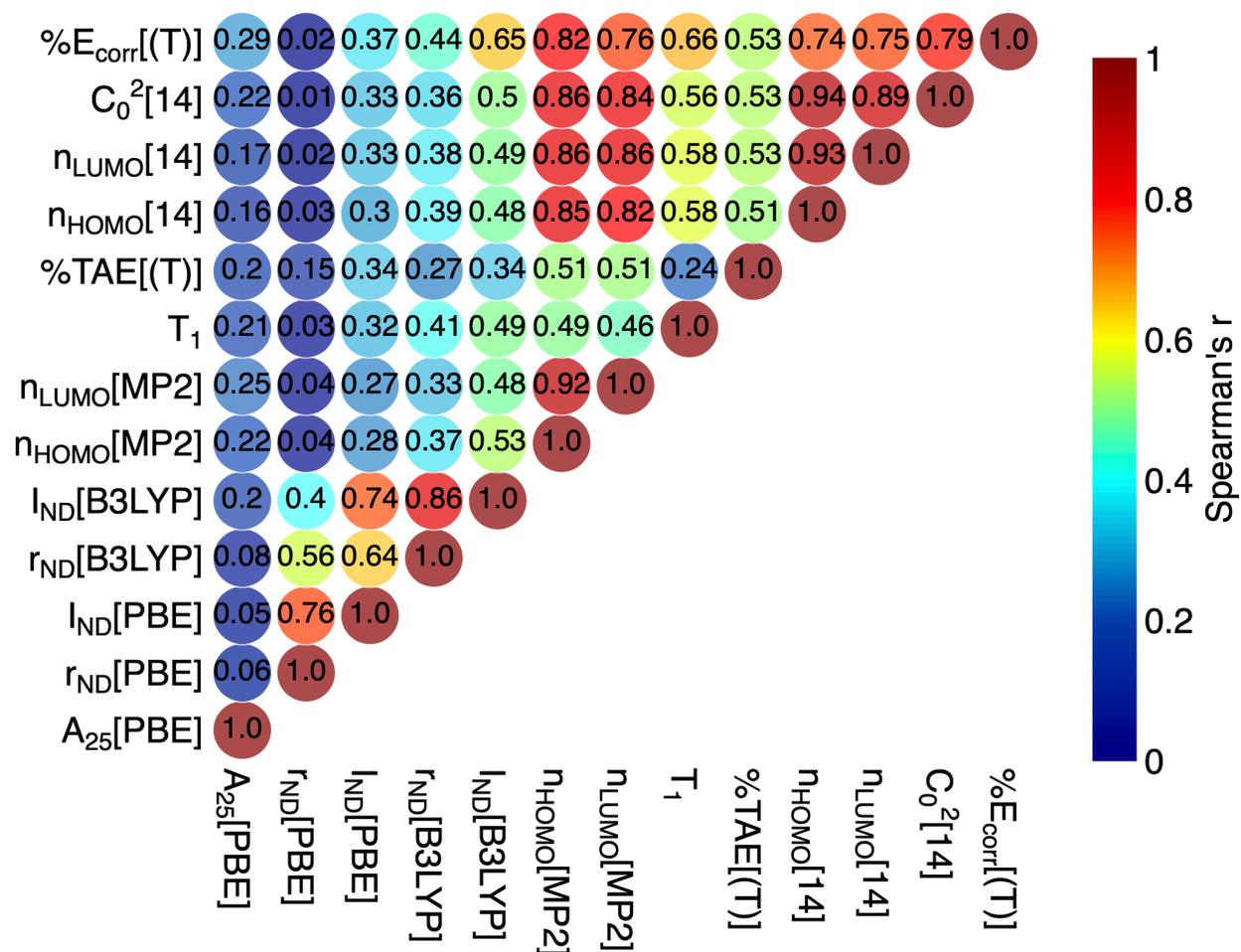

**Figure S5.** An upper triangular matrix of Spearman's *r* for pairs of MR diagnostics and %$E_{corr}$[(T)] on the set of 10,000 TMCs. For each pair, it is colored by the Spearman's *r* and the *r* value explicitly shown in the circle. Fourteen active orbitals are used for the CASSCF calculations

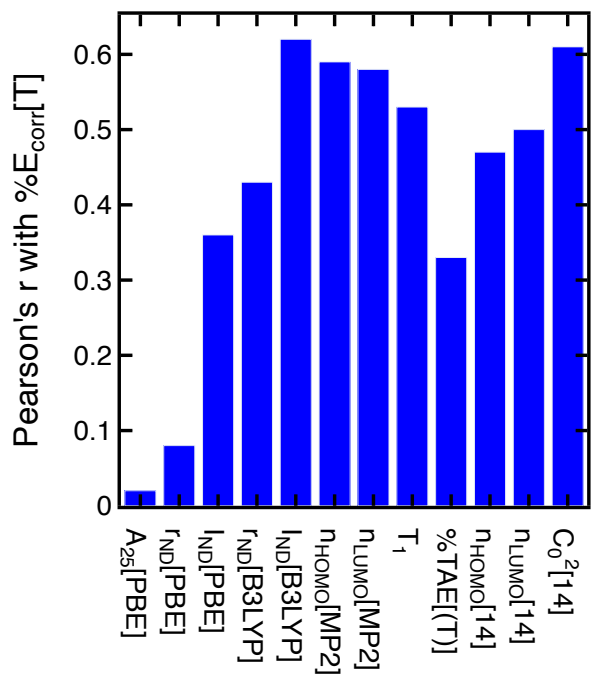

**Figure S6.** Bar plot of unsigned Pearson's *r* for %$E_{corr}$[(T)] with different MR diagnostics on the 10,000 TMCs. Fourteen active orbitals are used for the CASSCF calculations

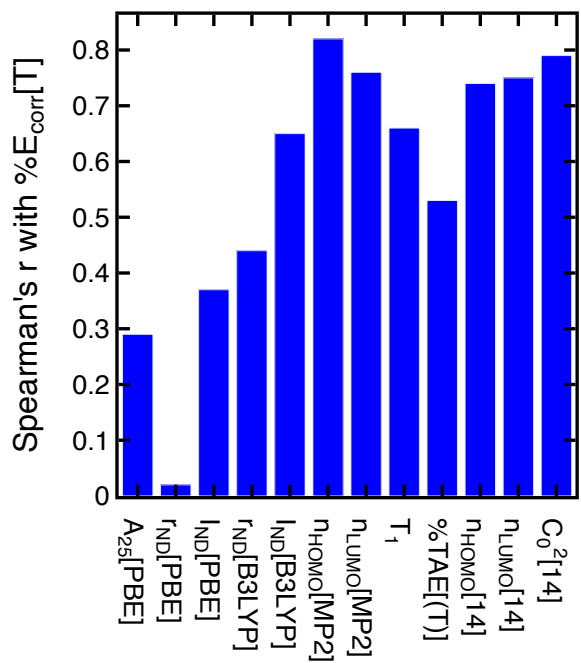

**Figure S7.** Bar plot of Spearman's *r* for %$E_{corr}$[(T)] with different MR diagnostics on the 10,000 TMCs. Fourteen active orbitals are used for the CASSCF calculations

Page S7

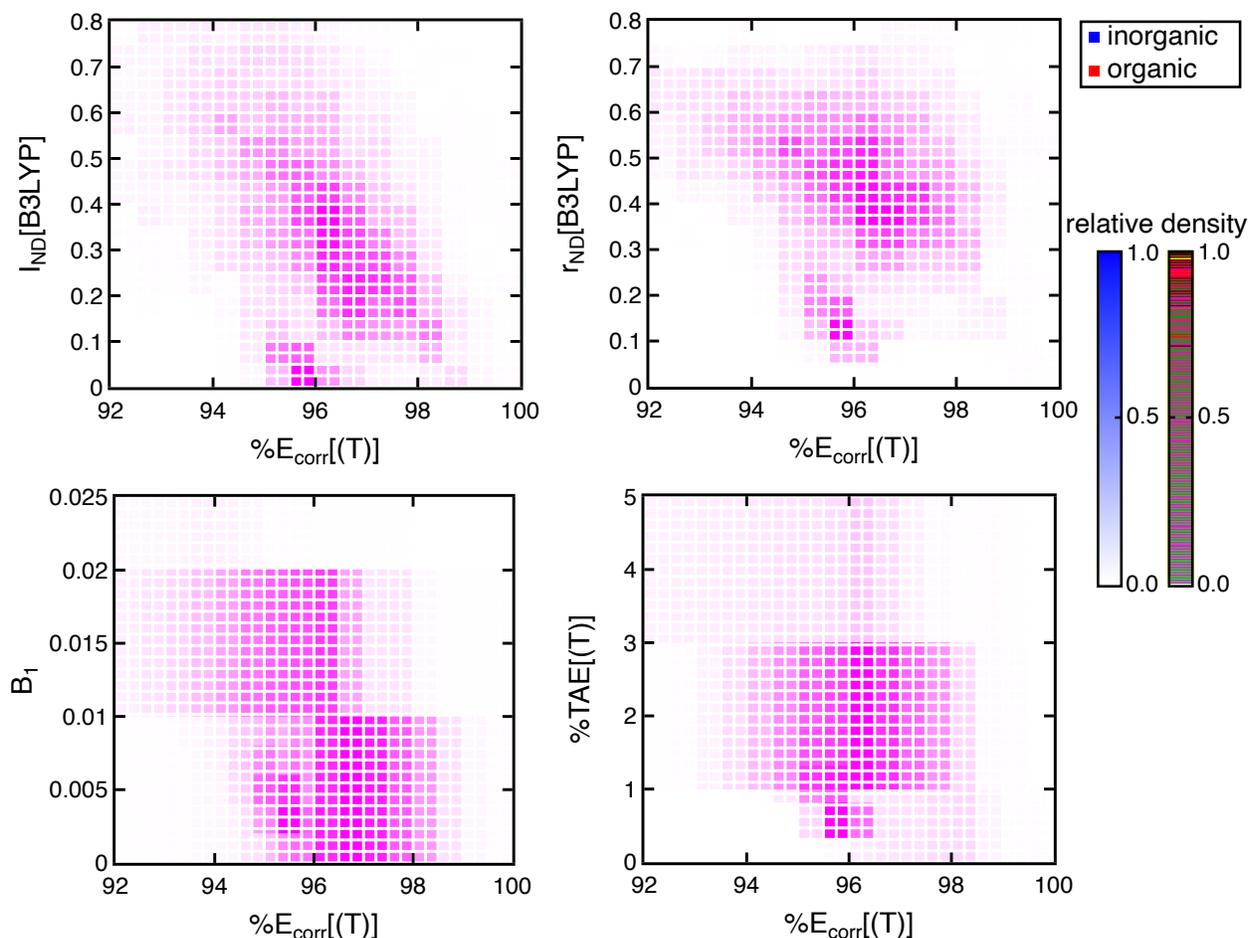

**Figure S8.** 2D histogram for %$E_{corr}$[(T)] vs. $I_{ND}$[B3LYP] (top left), $r_{ND}$[B3LYP] (top right), $B_1$ (bottom left), and %TAE[(T)] (bottom right) for the 10,000 TMCs in this work (blue) and for the 12,500 equilibrium or stretched organic molecules in our prior work[1] (AD-3165, PS-401, and LG-8934, red). The relative density of systems lying at a specific bin is represented by the opacity of the coloring, as shown for the color bars at right.

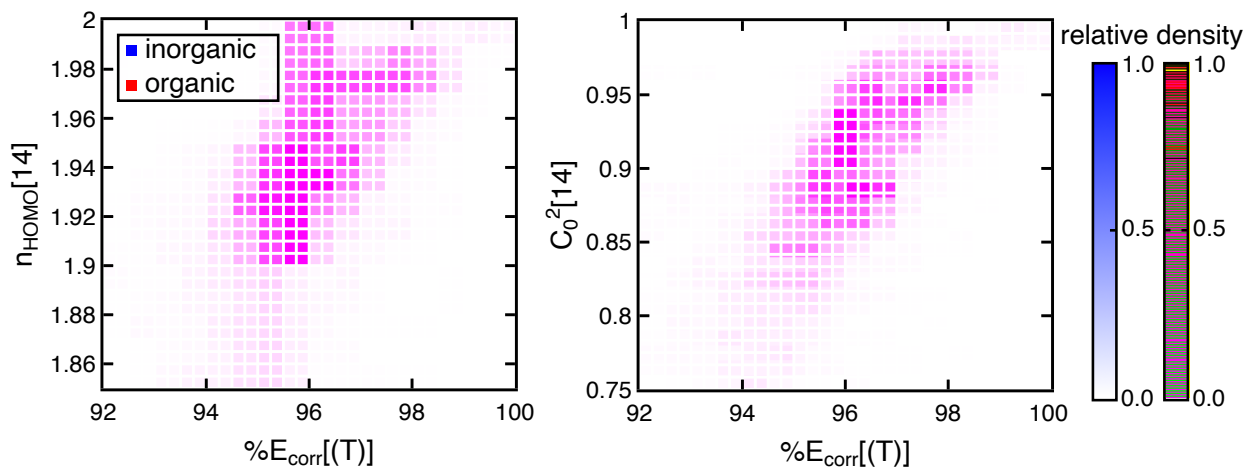

**Figure S9.** 2D histogram for %$E_{corr}$[(T)] vs. $n_{HOMO}$[14] (left) and $C_0^2$[14] (right) for the 10,000 TMCs in this work (blue) and for the 3566 equilibrium or stretched organic molecules in our prior work[1] (AD-3165 and PS-401, red). The relative density of systems lying at a specific bin is represented by the opacity of the coloring. Fourteen active orbitals are used for the CASSCF calculations. Note that LG-8934 is not included in the comparison since WFT-based diagnostics were not computed in that set.

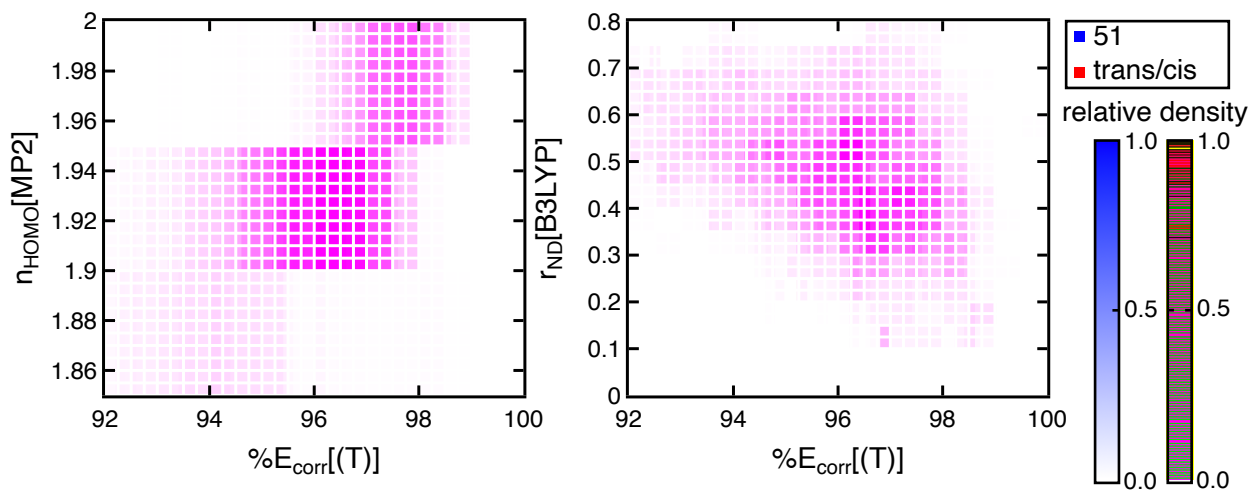

**Figure S10.** 2D histogram for %$E_{corr}$[(T)] vs. $n_{HOMO}$[MP2] (left) and $r_{ND}$[B3LYP] (right) for the 10,000 TMCs in this work. The 51 complexes (i.e., with only one non-He ligand) are shown in blue, and the *trans* and *cis* complexes (i.e., with two non-He ligands) are shown in red. The 51 complexes have a smaller overall size than the *trans* and *cis* complexes. The relative density of systems lying at a specific bin is represented by the opacity of the coloring.

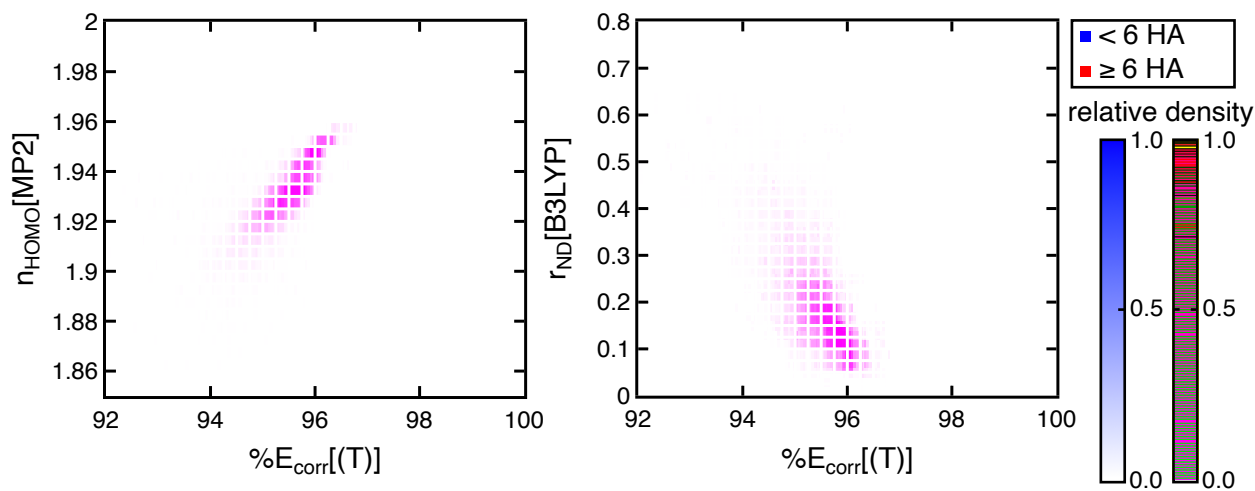

**Figure S11.** 2D histogram for %$E_{corr}$[(T)] vs. $n_{HOMO}$[MP2] (left) and $r_{ND}$[B3LYP] (right) for the 12,500 equilibrium or stretched organic molecules in our prior work[1], with the 3566 smaller molecules with < 6 heavy atoms (HA) shown in blue and 8934 larger molecules with ≥ 6 HAs shown in red. The relative density of systems lying at a specific bin is represented by the opacity of the coloring.

**Table S2.** Summary of ANN model performance on predicting WFT-based MR diagnostics and %$E_{corr}$[(T)] on the set-aside test set of 2,000 TMCs.

|  | MAE | scaled MAE | Pearson's r |
|---|---|---|---|
| $n_{HOMO}$[MP2] | 0.009 | 0.013 | 0.88 |
| $n_{LUMO}$[MP2] | 0.009 | 0.010 | 0.91 |
| $n_{HOMO}$[14] | 0.027 | 0.027 | 0.85 |
| $n_{LUMO}$[14] | 0.033 | 0.024 | 0.81 |
| $C_0^2$[14] | 0.019 | 0.023 | 0.90 |
| %TAE[(T)] | 0.348 | 0.007 | 0.90 |
| $T_1$ | 0.006 | 0.020 | 0.89 |
| max($t_1$) | 0.048 | 0.025 | 0.80 |
| %$E_{corr}$[(T)] | 0.211 | 0.016 | 0.94 |

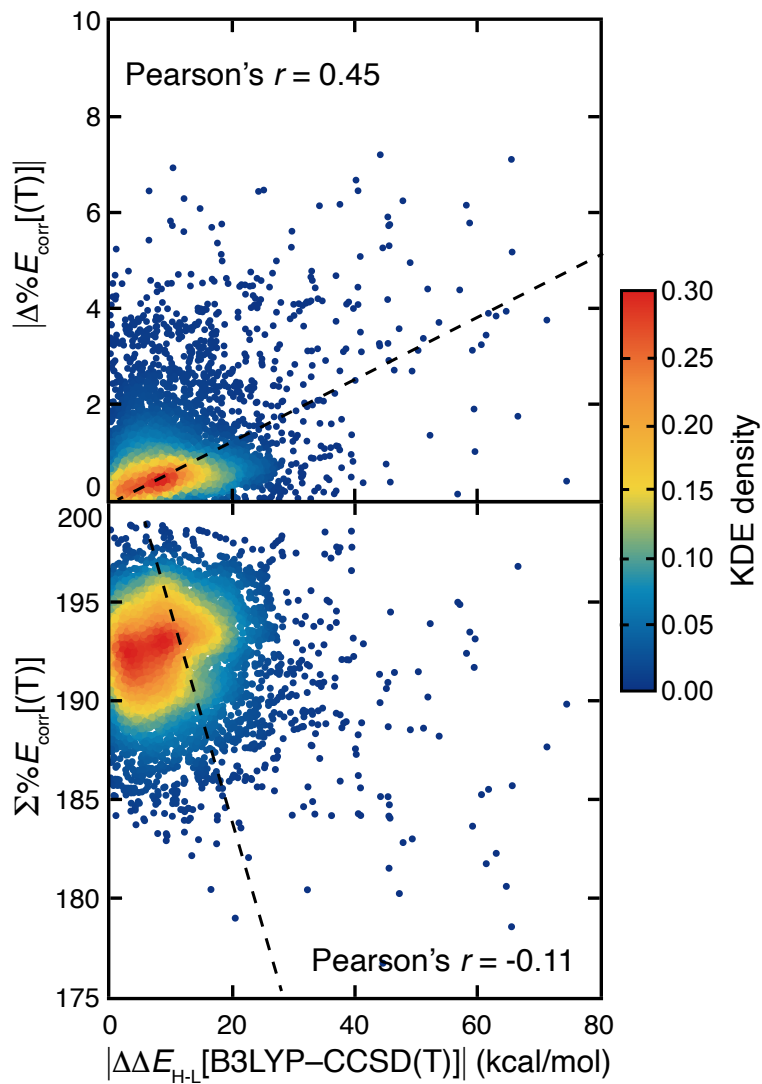

**Figure S12.** The absolute difference of adiabatic spin splitting between B3LYP and CCSD(T) (i.e., $|\Delta\Delta E_{\text{H-L}}[\text{B3LYP-CCSD(T)}]|$) vs. the absolute difference (top) and the sum (bottom) of $\%E_{\text{corr}}[(T)]$ of the two spin states, colored by kernel density estimation (KDE) density values, as indicated by inset color bars. A black dashed linear-regressed line is also shown in each case together with the Pearson correlation coefficients.

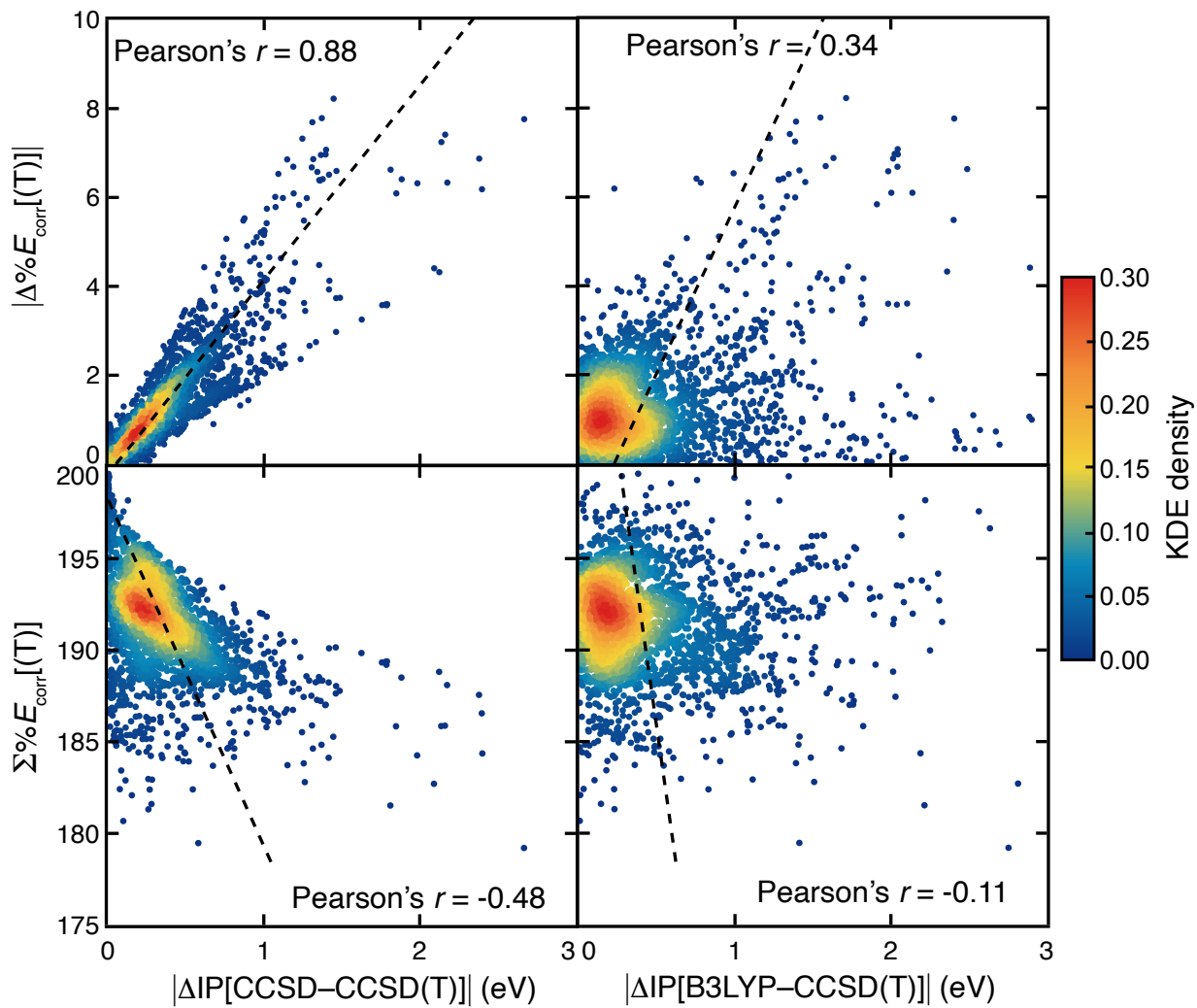

**Figure S13.** (left) The absolute difference of adiabatic IP between CCSD and CCSD(T) (i.e., |ΔIP[CCSD-CCSD(T)]|) vs. the absolute difference (top) and the sum (bottom) of %$E_{corr}$[(T)] of the two charge states. (right) The absolute difference of adiabatic IP between B3LYP and CCSD(T) (i.e., |ΔIP[B3LYP-CCSD(T)]|) vs. the absolute difference (top) and the sum (bottom) of %$E_{corr}$[(T)] of the two charge states. In both cases, points are colored by kernel density estimation (KDE) density values, as indicated by inset color bars. A black dashed linear-regressed line is also shown in each case together with the Pearson correlation coefficients.

**Table S3.** Features used for each prediction task.

| Target | Features |
|---|---|
| WFT-based diagnostics | CD-RACs, DFT-based diagnostics, oxidation state, spin state, and ligand charge |
| $\Delta\Delta E_{H-L}$ | CD-RACs from complexes at two spin states, oxidation state, and ligand charge, sums and differences of six DFT-based MR diagnostics, DFT evaluated $\Delta E_{H-L}$ with BLYP, B3LYP, PBE, and PBE0 |
| $\Delta$IP | CD-RACs of from complexes at two oxidation states, spin state of the ox-II complex, and ligand charge, sums and differences of six DFT-based MR diagnostics, DFT evaluated $\Delta E_{H-L}$ with BLYP, B3LYP, PBE, and PBE0 |

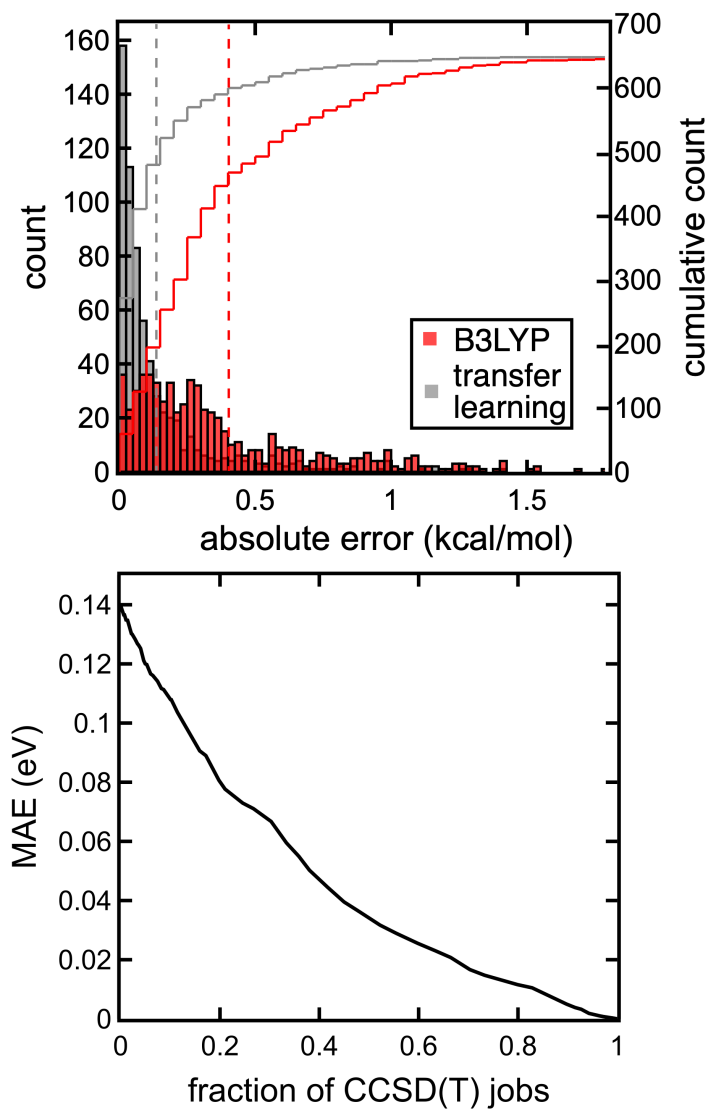

**Figure S14.** (top) Distributions of absolute errors for IP predicted DFT using B3LYP (red) and our transfer learning models (gray) on the set-aside test data, with the cumulative count shown according to the axis on the right. The MAEs are shown as vertical bars at 0.40 eV for DFT and 0.14 eV for transfer learning. (bottom) MAE of our multi-pronged strategy of transfer learning, uncertainty quantification, and multi-level modeling vs. the percentage of CCSD(T) calculations required. In both cases, we treat CCSD(T) results as our reference.

**Table S4.** MAE of $\Delta E_{H-L}$ and IP for B3LYP, MP2, CCSD, and their corresponding transfer learning approaches on the set-aside test data (1355 points for $\Delta E_{H-L}$ and 657 points for IP). In both cases, CCSD(T) is treated as the reference. The transfer learning approach leads to 5- to 10-fold reduction in MAE and systematically improves with the level of the quantum chemistry method. The inputs of the models are described in Table S9 and all the models as well as hyperparameters are included in the zip file of the Supporting Information.

|  | MAE of $\Delta E_{H-L}$ (kcal/mol) | MAE of IP (eV) |
|---|---|---|
| B3LYP | 10.2 | 0.40 |
| DFT-cost transfer learning | 2.8 | 0.14 |
| MP2 | 10.4 | 0.61 |
| MP2-cost transfer learning | 2.1 | 0.12 |
| CCSD | 5.6 | 0.35 |
| CCSD-cost transfer learning | 0.4 | 0.06 |

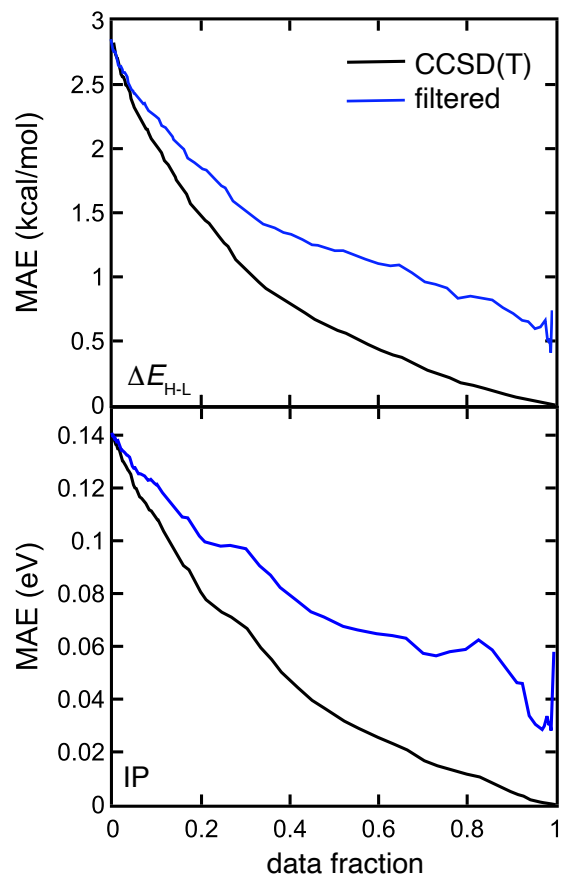

**Figure S15.** MAE of $\Delta E_{\text{H-L}}$ (top) and IP (bottom) for the multi-pronged strategy where the highest UQ points are explicitly computed by CCSD(T) (black) compared to those are directly filtered (i.e., eliminated) from the set-aside test data[15]. The x axis refers to the fraction of data considered to have high model uncertainty judged by a cutoff and are thus calculated or excluded. In the multi-pronged strategy, we perform CCSD(T) on this fraction of data and thus its error is counted as zero. In the direct filtering strategy, this fraction of data is directly removed from the test data. Note that if we randomly select points in the multi-pronged filtering strategy (instead of using UQ), the MAE would not decrease with respect to the data fraction (i.e., 2.8 kcal/mol $\Delta E_{\text{H-L}}$ and 1.4 eV for IP) but would have an error bar as the standard deviation of the absolute errors.

**Table S5.** Metals (M), oxidation states (ox), and spin states considered in this work. Cases where the high spin state is not calculated are shown with "--".

| $d$ electron configuration | M(ox) | High spin state | Intermediate spin state | Low spin state |
|---|---|---|---|---|
| $d^3$ | Cr(III) | -- | quartet | doublet |
| $d^4$ | Mn(III)/Cr(II) | quintet | triplet | singlet |
| $d^5$ | Fe(III)/Mn(II) | sextet | quartet | doublet |
| $d^6$ | Co(III)/Fe(II) | quintet | triplet | singlet |
| $d^7$ | Co(II) | -- | quartet | doublet |

**Table S6.** Summary of ligands studied. The ligands are either from the spectrochemical series (spectro) or our previous OHLDB set[16]. For each ligand, only the atoms with a non-zero form charges are shown for simplicity.

| Chemical name | Formula | SMILES string | Source | Formal charge | Connecting atom |
|---|---|---|---|---|---|
| amine | $NH_2^-$ | [NH2-] | spectro | [("N": -1)] | N |
| ammonia | $NH_3$ | [NH3] | spectro | [] | N |
| phosphide | $PH_2^-$ | [PH2-] | spectro | [("P": -1)] | P |
| phosphine | $PH_3$ | [PH3] | spectro | [] | P |
| azide | $N_3^-$ | [N-]=[N+]=[N-] | spectro | [("N": -1), ("N": 1), ("N": -1)] | N |
| carbonyl | CO | [C-]#[O+] | spectro | [("C": -1), ("O": 1)] | C |
| chloride | $Cl^-$ | [Cl-] | spectro | [("Cl": -1)] | Cl |
| flouride | $F^-$ | [F-] | spectro | [("F": -1)] | F |
| cyanide | $CN^-$ | [C-]#N | spectro | [("C": -1)] | S |
| hydrogen sulfide | $H_2S$ | [SH2] | spectro | [] | S |
| hydrosulfide | $HS^-$ | [HS-] | spectro | [("S": -1)] | O |
| water | $H_2O$ | [OH2] | spectro | [] | O |
| hydroxyl | $OH^-$ | [OH-] | spectro | [("O": -1)] | N |
| isothiocyanate | NCS- | [N-]=C=S | spectro | [("N": -1)] | S |
| thiocyante | SCN- | [S-]-C#N | spectro | [("S": -1)] | N |
| nitrito | $NO_2^-$ | [O-]-N=O | spectro | [("O": -1)] | N |
| C2 | $C_2$ | C4C | OHLDB | [] | C |
| nitrogen gas | $N_2$ | N#N | OHLDB | [] | N |
| oxygen gas | $O_2$ | O#O | OHLDB | [] | O |
| sulfidocarbon | CS | [C-]#[S+] | OHLDB | [("C": -1), ("S": 1)] | C |
| hydrogen cyanide | HCN | [CH]#N | OHLDB | [] | N |
| cyanate | NCO- | N#C[O-] | OHLDB | [("O": -1)] | N |
| oxoazanide | NO- | [N-]=O | OHLDB | [("N": -1)] | N |
| sulfanylideneazanide | NS- | [N-]=S | OHLDB | [("N": -1)] | N |
| formaldehyde | $H_2CO$ | O=[CH2] | OHLDB | [] | O |

**Table S7.** Type of TMCs and their DFT geometry optimization success rate. Metal-He distances are studied in 0.1 Å intervals from 1.4 to 1.9 Å and 0.2 Å intervals from 2.1 to 2.7 Å

| Type of TMCs | all-He | 1-non-He | trans | cis |
|---|---|---|---|---|
| Theoretical size | 220 | 5500 | 5500 | 5500 |
| Successful cases | 220 | 4321 | 4730 | 3577 |
| Success rate | 100% | 79% | 86% | 65% |

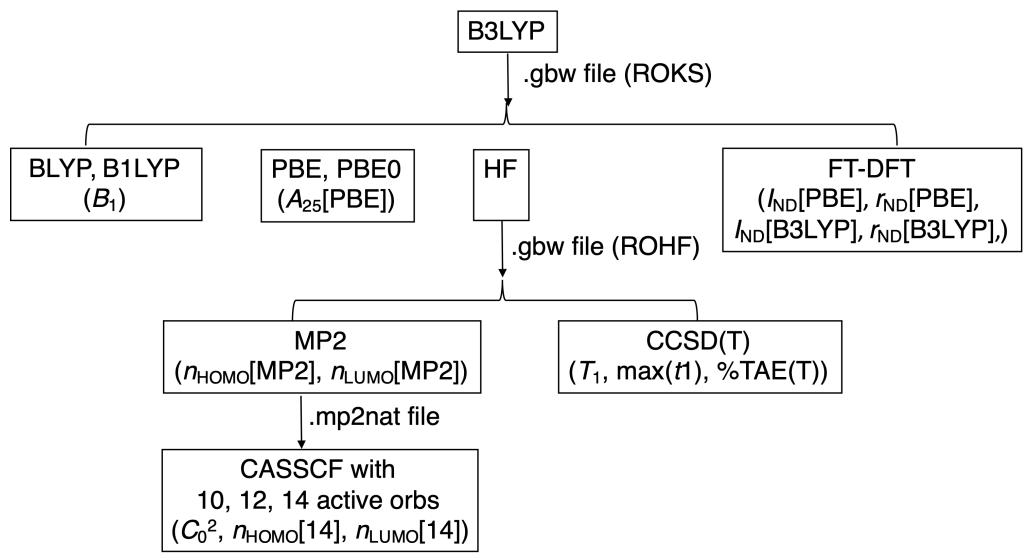

**Figure S16.** Workflow of computing 14 MR diagnostics. All calculations were performed with ORCA 4.0.2.1[17-18].

**Table S8.** Default convergence parameters used in self-consistent field calculations (i.e., HF, DFT, and CCSD(T)) for ORCA 4.0.2.1[17-18].

| Software | Energy convergence threshold (Ha) | DIIS error (Ha) | Maxiter |
|---|---|---|---|
| ORCA | 1e-6 | 1e-6 | 125 |

**Table S9.** MR diagnostics calculation attrition counts and reasons.

| Type | Count | Reason |
|---|---|---|
| Zero-temperature DFT (B3LYP, BLYP, B1LYP, PBE, and PBE0) | 983 | SCF convergence issue |
| Finite-temperature DFT (B3LYP and PBE) | 151 | SCF convergence issue |
| CASSCF at an active space of 14 orbitals | 424 | CASSCF convergence issue or exceeding the time limit of 48 hours. |
| CCSD(T) | 629 | MDCI convergence issue or exceeding the time limit of 48 hours. |

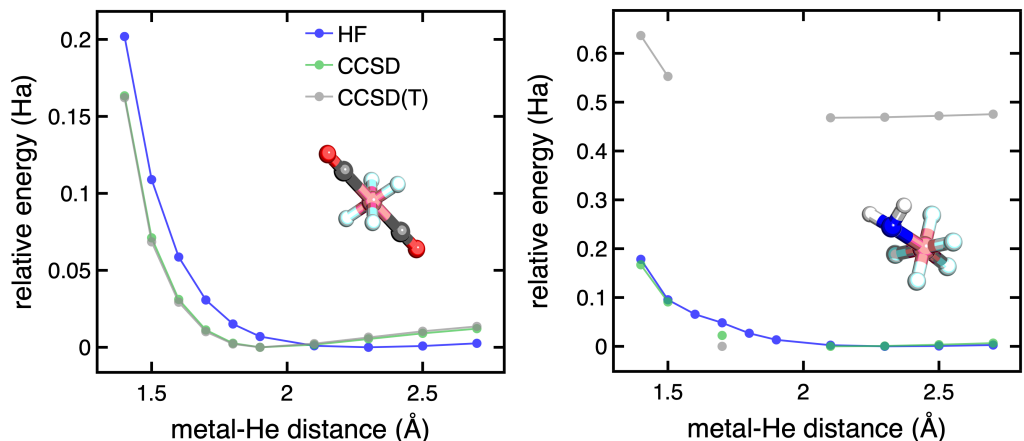

**Figure S17.** Potential energy curve for *trans* LS Co(II)He$_4$(CO)$_2$ (left) and LS Co(III)He$_5$(NH$_2^-$) (right) for HF (blue), CCSD (green), and CCSD(T) (gray) in Ha with increasing metal-He distance. For each method, the energies are shifted such that the minimum is set to 0 Ha. A discontinuity can be seen in LS Co(III)He$_5$(NH$_2^-$) at metal-He of 1.7 Å, resulted from the abnormally large perturbative correction at this metal-He distance.

**Table S10.** $T_1$ diagnostics for *trans* LS Co(II)He$_4$(CO)$_2$ (left) and LS Co(III)He$_5$(NH$_2^-$) with increasing metal-He distances. The $T_1$ diagnostic of *trans* LS Co(II)He$_4$(CO)$_2$ at metal-He distance of 1.7 Å is bolded. This discontinuity in $T_1$ diagnostic suggest the CCSD wavefunction was not converged to the same electronic structure at this particular metal-He distance. In cases where the CCSD calculation did not converge, the $T_1$ diagnostic is shown as "--".

| Metal-He distance (Å) | 1.4 | 1.5 | 1.6 | 1.7 | 1.8 | 1.9 | 2.1 | 2.3 | 2.5 | 2.7 |
|---|---|---|---|---|---|---|---|---|---|---|
| *trans* LS Co(II)He$_4$(CO)$_2$ | 0.05 | 0.08 | -- | **0.28** | -- | -- | 0.09 | 0.09 | 0.09 | 0.09 |
| LS Co(III)He$_5$(NH$_2^-$) | 0.02 | 0.03 | 0.03 | 0.03 | 0.03 | 0.03 | 0.03 | 0.03 | 0.02 | 0.02 |

**Table S11.** Cutoff values for the Grubbs test and Z-score test. A TMC was removed if it was marked as outlier by both the Grubbs and Z-score test.

| Grubbs | Z-score |
|---|---|
| 0.05 | 2.0 |

**Table S12.** Removed TMCs at each filtering step. The theoretical size of the dataset is 16,720.

| Reason | Number of points |
|---|---|
| Bad DFT optimized geometry | 3872 |
| Missing any of the 14 MR diagnostics | 2187 |
| Abnormal perturbative T correction | 274 |
| 14 orbitals not large enough as active space | 334 |

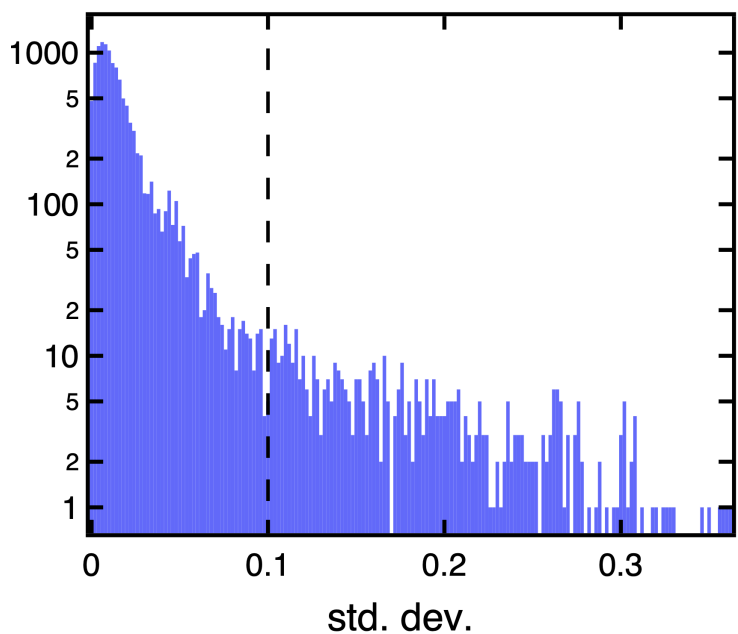

**Figure S18.** Distribution of the standard deviation (std. dev.) for $C_0^2$ diagnostic obtained by three active spaces (i.e., 10 orbitals, 12 orbitals, and 14 orbitals). The cutoff value of 0.1 is shown by a vertical dashed line. The *y* axis is in log scale.

**Text S1.** We have introduced a systematic approach to featurize molecular inorganic complexes that blends metal-centric and whole-complex topological properties in a feature set referred to as revised autocorrelation functions (RACs).[19] These RACs, variants of graph autocorrelations (ACs),[20-23] are sums of products and differences of atomic properties, i.e., electronegativity ($\chi$), nuclear charge (*Z*), topology (*T*), covalent radius (*S*), and identity (*I*). Standard ACs are defined as

$$P_d = \sum_i \sum_j P_i P_j \delta(d_{ij}, d)$$

where $P_d$ is the AC for property *P* at depth *d*, $\delta$ is the Dirac delta function, and $d_{ij}$ is the bond-wise path distance between atoms *i* and *j*.

In our approach, we have five types of RACs:
- $_{\text{all}}^{\text{f}}P_d$: standard ACs start on the full molecule (*f*) and have all atoms in the scope (all).
- $_{\text{ax}}^{\text{f}}P_d$ and $_{\text{eq}}^{\text{f}}P_d$: restricted-*scope* ACs that start on the full molecule (*f*) and separately evaluate axial or equatorial ligand properties

$$_{\text{ax/eq}}^{\text{f}}P_d = \frac{1}{|\text{ax/eq ligands}|} \sum_i^{n_{\text{ax/eq}}} \sum_i^{n_{\text{ax/eq}}} P_i P_j \delta(d_{ij}, d)$$

where $n_{\text{ax/eq}}$ is the number of atoms in the corresponding axial or equatorial ligand and properties are averaged within the ligand subtype.

- $^{mc}_{all}P_d$: restricted-scope, metal-centered (mc) descriptors that start on the metal center (mc) and have all atoms in the scope (all), in which one of the atoms, $i$, in the $i,j$ pair is a metal center:

$$^{mc}_{all}P_d = \sum_i^{mc}\sum_i^{all} P_i P_j \delta(d_{ij}, d)$$

- $^{lc}_{ax}P_d$ :and $^{lc}_{ax}P_d$: restricted-scope, metal-proximal ACs that start on a ligand-centered (lc) and separately evaluate axial or equatorial ligand properties, in which one of the atoms, $i$, in the $i,j$ pair is the metal-coordinating atom of the ligand:

$$^{lc}_{ax/eq}P_d = \frac{1}{|ax/eq\ ligands|}\frac{1}{|lc|}\sum_i^{lc}\sum_i^{n_{ax/eq}} P_i P_j \delta(d_{ij}, d)$$

We also modify the AC definition, $P'$, to property differences rather than products for a minimum depth, d, of 1 (as the depth-0 differences are always zero):

$$^{lc/mc}_{ax/eq/all}P'_d = \sum_i^{lc\ or\ mc}\sum_i^{scope} P_i P_j \delta(d_{ij}, d)$$

where scope can be axial, equatorial, or all ligands.

Although RACs have been demonstrated to be accurate for predicting equilibrium properties with ML models, including spin-splitting energies and redox potential[24-25], they encode no explicit 3D geometry information and cannot distinguish distorted geometries from equilibrium structures. Therefore, we combine RACs and a 3D geometry based representation, Coulomb matrix (CM), in a new representation we refer to as the Coulomb-decay revised autocorrelations (CD-RACs)[1].

$$P_{d,CD} = \frac{1}{n}\begin{cases} \sum_i^n\sum_j^n \frac{P_i P_j}{r_{ij}}\delta(d_{ij}, d), & d > 0 \\ \frac{1}{2}\sum_i^n P_i^{2.4}, & d = 0 \end{cases}$$

The form of CD-RACs is in analog of RACs, but is simply scaled by the pairwise atom distance when the bond depth is not zero. When the bond depth is zero, we use the power of 2.4 and introduce a pre-factor of 0.5 as in CM.

We calculate CD-RACs at all starting points (lc, mc, and f), all scopes (eq, ax, and all), and consider both the products and differences, which gives 180 CD-RACs in total. Among the CD-RACs, some are constant either due to their nature (e.g., full-complex depth-0 $I$ CD-RAC is always 0.5) or small sizes of TMCs in the dataset (e.g., all full-complex depth-3 CD-RACs are 0). Eliminating those constant CD-RACs yields 134 CD-RACs in total.

**Table S13**. Range of hyperparameters sampled for ANN models trained from scratch with Hyperopt[26]. The lists in the architecture row can refer to two or three hidden layers (i.e., the number of items in the list), and the number of nodes in each layer are denoted as elements of the list. The built-in Tree of Parzen Estimator algorithm in Hyperopt was used for the hyperparameter selection process.

| Architecture | ([128], [256], [512], [128, 128], [256, 256], [512, 512], [128, 128, 128], [256, 256, 256], [512, 512, 512]) |
|---|---|
| L2 regularization | [1e-6, 1] |
| Dropout rate | [0, 0.5] |
| Learning rate | [1e-5, 1e-3] |
| Beta1 | [0.75, 0.99] |
| Batch size | [16, 32, 64, 128, 256, 512] |

# References

1. Duan, C.; Liu, F.; Nandy, A.; Kulik, H. J., Data-Driven Approaches Can Overcome the Cost– Accuracy Trade-off in Multireference Diagnostics. *J Chem Theory Comput* **2020,** https://dx.doi.org/10.1021/acs.jctc.0c00358 (16), 4373-4387.
2. Schultz, N. E.; Zhao, Y.; Truhlar, D. G., Density functionals for inorganometallic and organometallic chemistry. *Journal of Physical Chemistry A* **2005,** *109* (49), 11127-11143.
3. Fogueri, U. R.; Kozuch, S.; Karton, A.; Martin, J. M. L., A simple DFT-based diagnostic for nondynamical correlation. *Theoretical Chemistry Accounts* **2013,** *132* (1), 1291.
4. Ramos-Cordoba, E.; Salvador, P.; Matito, E., Separation of dynamic and nondynamic correlation. *Physical Chemistry Chemical Physics* **2016,** *18* (34), 24015-24023.
5. Ramos-Cordoba, E.; Matito, E., Local Descriptors of Dynamic and Nondynamic Correlation. *J Chem Theory Comput* **2017,** *13* (6), 2705-2711.
6. Kesharwani, M. K.; Sylvetsky, N.; Kohn, A.; Tew, D. P.; Martin, J. M. L., Do CCSD and approximate CCSD-F12 variants converge to the same basis set limits? The case of atomization energies. *J Chem Phys* **2018,** *149* (15), 154109.
7. Jensen, H. J. A.; Jorgensen, P.; Ågren, H.; Olsen, J., Second-order Moller–Plesset perturbation theory as a configuration and orbital generator in multiconfiguration self-consistent field calculations. *The Journal of Chemical Physics* **1988,** *88* (6), 3834-3839.
8. Lee, T. J.; Taylor, P. R., A Diagnostic for Determining the Quality of Single-Reference Electron Correlation Methods. *Int J Quantum Chem* **1989**, 199-207.
9. Janssen, C. L.; Nielsen, I. M. B., New diagnostics for coupled-cluster and Moller-Plesset perturbation theory. *Chemical Physics Letters* **1998,** *290* (4-6), 423-430.
10. Karton, A.; Daon, S.; Martin, J. M. L., W4-11: A high-confidence benchmark dataset for computational thermochemistry derived from first-principles W4 data. *Chemical Physics Letters* **2011,** *510* (4-6), 165-178.
11. Sears, J. S.; Sherrill, C. D., Assessing the performance of density functional theory for the electronic structure of metal-salens: The d(2)-metals. *Journal of Physical Chemistry A* **2008,** *112* (29), 6741-6752.
12. Sears, J. S.; Sherrill, C. D., Assessing the performance of density functional theory for the electronic structure of metal-salens: The 3d(0)-metals. *Journal of Physical Chemistry A* **2008,** *112* (15), 3466-3477.
13. Langhoff, S. R.; Davidson, E. R., Configuration interaction calculations on the nitrogen molecule. *Int J Quantum Chem* **1974,** *8* (1), 61-72.
14. Tishchenko, O.; Zheng, J. J.; Truhlar, D. G., Multireference model chemistries for thermochemical kinetics. *J Chem Theory Comput* **2008,** *4* (8), 1208-1219.
15. Janet, J. P.; Duan, C.; Yang, T.; Nandy, A.; Kulik, H. J., A quantitative uncertainty metric controls error in neural network-driven chemical discovery. *Chemical Science* **2019**.
16. Gugler, S.; Janet, J. P.; Kulik, H. J., Enumeration of de novo inorganic complexes for chemical discovery and machine learning. *Mol Syst Des Eng* **2020,** *5* (1), 139-152.
17. Neese, F., The ORCA program system. *Wires Comput Mol Sci* **2012,** *2* (1), 73-78.
18. Neese, F., Software update: the ORCA program system, version 4.0. *Wires Comput Mol Sci* **2018,** *8* (1), e1327.
19. Janet, J. P.; Kulik, H. J., Resolving transition metal chemical space: feature selection for machine learning and structure-property relationships. *Journal of Physical Chemistry A* **2017,** *121* (46), 8939-8954.

20. Devillers, J.; Domine, D.; Guillon, C.; Bintein, S.; Karcher, W., Prediction of partition coefficients (log p oct) using autocorrelation descriptors. *SAR QSAR Environ. Res.* **1997,** *7* (1-4), 151-172.
21. Broto, P.; Devillers, J., *Autocorrelation of properties distributed on molecular graphs*. Kluwer Academic Publishers: Dordrecht, The Netherlands: 1990.
22. Virshup, A. M.; Contreras-García, J.; Wipf, P.; Yang, W.; Beratan, D. N., Stochastic voyages into uncharted chemical space produce a representative library of all possible drug-like compounds. *Journal of the American Chemical Society* **2013,** *135* (19), 7296-7303.
23. Broto, P.; Moreau, G.; Vandycke, C., Molecular structures: perception, autocorrelation descriptor and sar studies: system of atomic contributions for the calculation of the n-octanol/water partition coefficients. *European journal of medicinal chemistry* **1984,** *19* (1), 71-78.
24. Janet, J. P.; Ramesh, S.; Duan, C.; Kulik, H. J., Accurate multi-objective design in a space of millions of transition metal complexes with neural-network-driven efficient global optimization. *ACS Central Science* **2020,** *6* (4), 513-524.
25. Nandy, A.; Duan, C.; Janet, J. P.; Gugler, S.; Kulik, H. J., Strategies and Software for Machine Learning Accelerated Discovery in Transition Metal Chemistry. *Industrial & Engineering Chemistry Research* **2018,** *57* (42), 13973-13986.
26. Bergstra, J.; Yamins, D.; Cox, D. D. In *Hyperopt: A python library for optimizing the hyperparameters of machine learning algorithms*, Proceedings of the 12th Python in science conference, 2013; pp 13-20.